# A new approach to solving the radiation field problem of an extended helical undulator

M. I. Ivanyan[1], B. Grigoryan[1], A. Grigoryan[1, 2], L. Aslyan[1], V. Avagyan[1, 3], H. Babujyan[1], S. Arutunian[1, 3], K. Floettmann[4], F. Lemery[4]

[1]*CANDLE SRI, Acharyan 31, 0022, Yerevan, Armenia*

[2]*Yerevan State University, Yerevan, Armenia*

[3]A. Alikhanian National Scientific Laboratory, Yerevan, Armenia

[4]Deutsches Elektronen-Synchrotron DESY, Notkestraße 85, 22607 Hamburg, Germany

*Abstract*

A new method is applied to construct an exact solution for the radiation field of a particle moving along an infinite helical trajectory. The solution is obtained in the form of a series expansion in cylindrical multipoles. The obtained solution is compared with the existing approximate solution and, using the derived exact relationships for the Doppler effect, is used to construct integral and angular characteristics of the radiation field. The possibility of a continuous transition from expressions for a helical trajectory of a particle to expressions describing the motion of a particle along a closed circle is shown. Optimization criteria are introduced and the possibility of optimizing the radiation characteristics by several parameters is considered.

## 1. INTRODUCTION

Helical undulators are periodic magnetic structures which force charged particles – in general electrons – onto helical trajectories, whereby they emit strong synchrotron radiation in forward direction [1 – 6]. In contrast to radiation produced in planar undulators, the generated on-axis radiation field is circularly polarized in case of helical particle trajectories, which makes helical undulators to particularly interesting devices. The standard approach to calculate the radiation properties of undulators involves integration of retarded Lienhard-Wiechert potentials and some simplifying assumptions, like relativistic particle energies ($\gamma \gg 1$) and an observation point in the far zone, so that a small angle approximation ($\sin\theta \cong \theta$) can be applied for the observation angle of the radiation $\theta$ [5], see Figure 1, left.

In the following, a new approach is developed, leading to a solution for the radiation field, which is valid at any point in space, inside and outside the particle trajectory, and at any particle energy. Especially at low particles energies, as they are relevant, e.g., for some THz sources, deviations of the exact (undisturbed) radiation spectrum from the approximate solution can be found, as will be discussed in section 5.

The approach can be generalized to include, e.g., a resistive vacuum pipe [7].

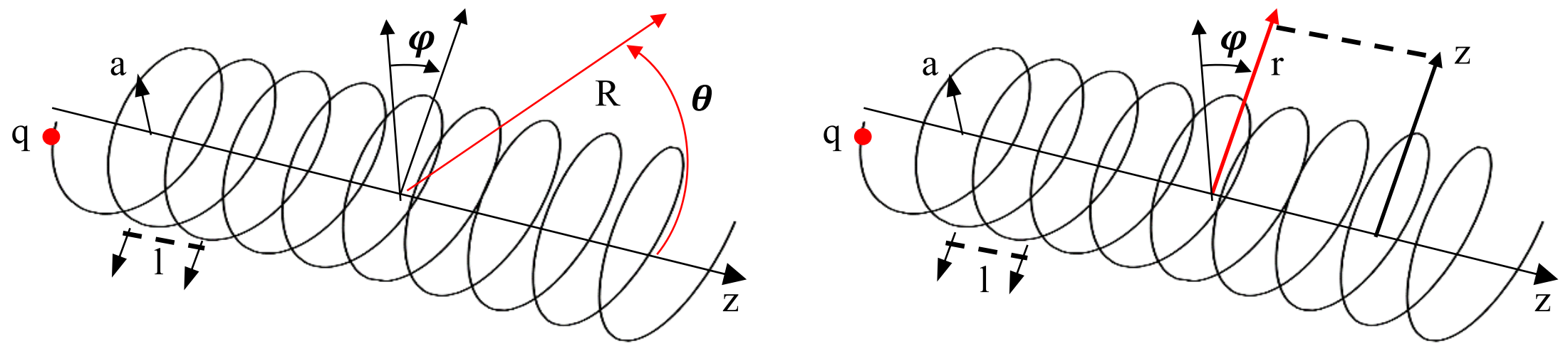

Figure 1: Coordinate systems for the description of helical trajectories and radiation fields of undulators. Left: spherical coordinate system commonly used in calculations; right: cylindrical coordinate system used in this work.

Since the particle trajectory has a cylindrical symmetry, a cylindrical coordinate system $r, \varphi, z$, aligned to the axis of the spiral, is used (Fig. 1, right). The period of the electrons spiral trajectory is $l$, its radius is $a$, the Lorentz factor of the particle is $\gamma$, its longitudinal velocity – parallel to the spiral axis – is $v_z$, and the rotation frequency around the spiral axis $\omega_0$. All quantities are constants of motion and the spiral trajectory is assumed to be infinitely long, so that the solution describes the steady state (as it is the case for the standard solution). The extension of the presented approach to the transition into or out of an undulator is discussed in [8].

With these assumptions a uniform description of the fields in the near and far zones, both outside the cylindrical surface $r = a$ containing the particle trajectory, and in the region $r < a$ located inside this surface, can be formulated by applying the partial domain method.

The search form of the fields, in the external region ($r > a$) and the internal region ($r < a$) of space, which are separated by the surface $r = a$, needs to be represented differently: in the first case it must satisfy the condition of the radiation field at infinity, and in the second case it must not contain divergences on the spiral axis ($r = 0$).

Different search forms of the fields in the two distinct areas of space lead to a discontinuity of the field values at the interface $r = a$, which is natural, because this surface contains charges and currents due to the particle trajectory. The connection of the field components on both sides of the interface and the charges and currents located on it, is reached through the boundary conditions as they arise from the integral form of Maxwell's equations [9].

## 2. FIELD REPRESENTATION BY SUPERPOSITION OF TM AND TE MODES

The search form of the field components, which are elementary solutions of the inhomogeneous Maxwell equations in the frequency domain, are initially represented in the form of a multipole expansion with discontinuous functions on the cylindrical surface $r = a$ , i.e., coaxial to the axis of the spiral trajectory.

For the case of a particle motion in free space, it has the form [10]:

$$\vec{\mathcal{E}} = \begin{cases} \sum_{m=-\infty}^{\infty} \vec{\mathcal{E}}_m^{(J)} \\ \sum_{m=-\infty}^{\infty} \vec{\mathcal{E}}_m^{(H)} \end{cases}, \qquad \vec{\mathcal{H}} = \begin{cases} \sum_{m=-\infty}^{\infty} \vec{\mathcal{H}}_m^{(J)} & r < a \\ \sum_{m=-\infty}^{\infty} \vec{\mathcal{H}}_m^{(H)} & r > a \end{cases}. \tag{1}$$

Each term of the multipole expansion is represented as a superposition of TM $\left(\vec{E}_{m,TM}^{(Z)}\right)$ and TE $\left(\vec{E}_{m,TE}^{(Z)}\right)$ modes, with $Z = J$ or $H$, and arbitrary weight factors $\left(\mathcal{A}_m^{(Z)}, \mathcal{B}_m^{(Z)}\right)$, i. e. it is composed of fundamental solutions of the homogeneous Maxwell equations in cylindrical coordinates:

$$\begin{aligned} \vec{\mathcal{E}}_m^{(Z)} &= \mathcal{A}_m^{(Z)} \vec{E}_{m,TM}^{(Z)} + \mathcal{B}_m^{(Z)} \vec{E}_{m,TE}^{(Z)} \\ Z_0 \vec{\mathcal{H}}_m^{(Z)} &= \mathcal{A}_m^{(Z)} \vec{H}_{m,TM}^{(Z)} + \mathcal{B}_m^{(Z)} \vec{H}_{m,TE}^{(Z)} \end{aligned}, \tag{2}$$

where

$$\begin{aligned} \vec{E}_{m,TM}^{(Z)} &= -\nu_m^{-2} \operatorname{rot} \vec{R}, \quad Z_0 \vec{H}_{m,TM}^{(Z)} = jk\nu_m^{-2} \vec{R}, \\ Z_0 \vec{H}_{m,TE}^{(Z)} &= -\nu_m^{-2} \operatorname{rot} \vec{R}, \quad \vec{E}_{m,TE}^{(,Z)} = -jk\nu_m^{-2} \vec{R}\,. \\ \vec{R} &= \vec{e}_z \times \vec{\nabla}\left(Z_m e^{j(m\varphi + pz - \omega t)}\right), \quad k = \omega/c \end{aligned} \tag{3}$$

If $Z = J$, i.e. $r < a$, $Z_m = J_m(\nu_m r)$ and if $Z = H$, i.e. $r > a$, $Z_m = H_m^{(1)}(\nu_m r)$, where $J_m$ and $H_m^{(1)}$ are the Bessel function and the Hankel function of the first kind; $Z_0 = (\varepsilon_0 c)^{-1}$ is the impedance of free space, $\varepsilon_0$ is the dielectric constant of vacuum, $c$ is the speed of light, and $\omega$ is the rotational frequency of the particle. In Eqn. 3 $p_m$ and $\nu_m = \sqrt{\omega^2/c^2 - p_m^2}$ are the longitudinal and transverse eigenvalues of the $m^{\text{th}}$ mode. In the case of a linear motion of the particle $p_m = \omega/\text{v}$ (v is total velocity of the particle) and $\nu_m = j\omega/\text{v}\gamma$ ($j$ is the imaginary unit), while for the helical motion $p_m = (\omega - m\omega_0)/\text{v}_z$ ($\text{v}_z$ is longitudinal component of the particle velocity) and

$$\nu_m = \sqrt{\omega^2/c^2 - (\omega - m\omega_0)^2/\text{v}_z^2}. \tag{4}$$

In the spatiotemporal representation, the density distribution (surface density) of the charge of a point particle (with charge $q$) moving along a helical trajectory located on the surface $r = a$ has the form:

$$\rho = q\delta(\varphi - \omega_0 t)\delta(z - v_z t) \tag{5}$$

Its corresponding spectral decomposition is written as follows:

$$\tilde{\rho} = q \sum_{m=1}^{\infty} \chi_m e^{jm\varphi} e^{j\{pz - (m\omega_0 + pv_z)t\}} \tag{6}$$

The expression $m\omega_0 + pv_z$ standing before the time $t$ in the exponential in (5) has the meaning of frequency, which should be common to all terms of the expansion, i.e., independent of the index m and should be relabelled: $\omega = m\omega_0 + pv_z$,. The longitudinal wave number $p$, on the contrary, depends on $m$ and is expressed in terms of frequency as follows: $p = p_m = (\omega - m\omega_0)/v_z$.

Let us recall that we introduced the same notation for $p$ when writing down the elements of field decomposition (1-3). This ensures coordination of the phases of multipole expansions of fields with the phases of expansions (6) of charges.

Thus, the density distribution of a point charge moving along a helical trajectory can be represented as the following expansion:

$$\tilde{\rho} = \sum_{m=1}^{\infty} \chi_m \tilde{\rho}_m \tag{7}$$

where

$$\tilde{\rho}_m = q e^{jm\varphi} e^{j\{p_m z - \omega t\}} \tag{8}$$

and $\chi_m$ is a proportionality factor which needs to be determined. It has the meaning of an expansion coefficient of the charge and current densities into multipoles. In the case of a linear motion, it does not depend on m and is proportional to $a^{-1}$ [11,12], but in the case of a helical motion, its definition requires additional considerations.

To determine the amplitudes $\mathcal{A}_m^{(J)}, \mathcal{A}_m^{(H)}, \mathcal{B}_m^{(J)}, \mathcal{B}_m^{(H)}$, boundary conditions [9] are used, so that a connection of the terms of field expansion into multipoles on both sides of the surface $r = a$, with the charge $\tilde{\rho}_m$ and current $\vec{\tilde{j}}_m = \tilde{\rho}_m\{0,\ \omega_0 a,\ \mathrm{v}_z\}$ densities expansion terms are established. The discontinuities of the tangential magnetic components $\mathcal{H}_{m,\varphi}^{(Z)}$ and $\mathcal{H}_{m,z}^{(Z)}$ are proportional to the tangential components $\tilde{j}_{m,z}$ and $\tilde{j}_{m,\varphi}$ of the current density and normal to them, i.e., $\mathcal{H}_{m,\varphi}^{(Z)} \propto \tilde{j}_{m,z}$ and $\mathcal{H}_{m,z}^{(Z)} \propto \tilde{j}_{m,\varphi}$ . The discontinuity of the normal electrical component $\mathcal{E}_{m,r}^{(Z)}$ is proportional to the charge density $\tilde{\rho}_m$. The tangential electric components $\mathcal{E}_{m,\varphi}^{(Z)}$, $\mathcal{E}_{m,z}^{(Z)}$ and the normal magnetic component $\mathcal{H}_{m,r}^{(Z)}$ are continuous on the specified surface.

In the common cases of rectilinear or helical motion, the above conditions can be represented by a system of four vector equations:

$$\begin{gathered}
\left(\vec{\mathcal{E}}_m^{(H)}/_{r\to a+} - \vec{\mathcal{E}}_m^{(J)}/_{r\to a-}\right) \times \vec{e}_r = 0 \\
\left(\vec{\mathcal{H}}_m^{(H)}/_{r\to a+} - \vec{\mathcal{H}}_m^{(J)}/_{r\to a-}\right) \times \vec{e}_r = \chi_m \vec{\tilde{j}}_m \\
\left(\vec{\mathcal{E}}_m^{(H)}/_{r\to a+} - \vec{\mathcal{E}}_m^{(J)}/_{r\to a-}\right) \cdot \vec{e}_r = -\chi_m \tilde{\rho}_m/\varepsilon_0 \\
\left(\vec{\mathcal{H}}_m^{(H)}/_{r\to a+} - \vec{\mathcal{H}}_m^{(J)}/_{r\to a-}\right) \cdot \vec{e}_r = 0.
\end{gathered} \tag{9}$$

Here $\vec{e}_r$ is the radial unit vector (perpendicular to the surface $r = a$).

In the case of a helical motion, i.e., $j_\varphi \neq 0$, Eqn. 9 is reduced to a system of four linearly independent equations, with the same number of unknowns. For $r = a$ we have:

$$\begin{gathered}
-\mathcal{A}_m^{(H)} \vec{H}_{m,TM_\varphi}^{(H)} + \mathcal{A}_m^{(J)} \vec{H}_{m,TM_\varphi}^{(J)} = \chi_m \tilde{\rho}_m (V - m p_m \omega_0/\nu_m^2) \\
\mathcal{A}_m^{(H)} \vec{E}_{m,TM_r}^{(H)} - \mathcal{A}_m^{(J)} \vec{E}_{m,TM_r}^{(J)} + \mathcal{B}_m^{(H)} \vec{E}_{m,TE_r}^{(H)} - \mathcal{B}_m^{(J)} \vec{E}_{m,TE_r}^{(J)} = -\chi_m \tilde{\rho}_m/\varepsilon_0 \\
\mathcal{A}_m^{(H)} \vec{H}_{m,TM_r}^{(H)} - \mathcal{A}_m^{(J)} \vec{H}_{m,TM_r}^{(J)} + \mathcal{B}_m^{(H)} \vec{H}_{m,TE_r}^{(H)} - \mathcal{B}_m^{(J)} \vec{H}_{m,TE_r}^{(J)} = 0 \\
-\mathcal{B}_m^{(H)} \vec{E}_{m,TE_\varphi}^{(H)} + \mathcal{B}_m^{(J)} \vec{E}_{m,TE_\varphi}^{(J)} = 0\ .
\end{gathered} \tag{10}$$

The determinant of this system is equal to $4m^2\omega^4/\pi^2 c^4 a^4 \nu_m^8$. It does not vanish when $m > 0$ and is not equal to infinity when $\nu_m \neq 0$. For these cases, the amplitudes are determined using Cramer's rule:

$$\begin{matrix} \mathcal{A}_m^{(J)} \\ \mathcal{A}_m^{(H)} \end{matrix} = q \frac{\pi a}{2\varepsilon_0 V \omega} \chi_m f_m \begin{Bmatrix} H_m^{(1)}(a\nu_m) \\ J_m(a\nu_m) \end{Bmatrix}, \quad \begin{matrix} \mathcal{B}_m^{(J)} \\ \mathcal{B}_m^{(H)} \end{matrix} = jq \frac{\pi a^2 \omega_0}{2\varepsilon_0 c} \chi_m \nu_m \begin{Bmatrix} H'^{(1)}_m(a\nu_m) \\ J'_m(a\nu_m) \end{Bmatrix} \tag{11}$$

$$f_m = \omega(\omega_0 m - \omega/\gamma_z^2),\ \ \gamma_z^2 = (1 - \mathrm{v}_z^2/c^2)^{-1}. \tag{12}$$

$J'_m$ and $H'^{(1)}_m$ are the derivatives of the Bessel function and the Hankel function of the first kind.

## 3. MUTUAL COUPLING OF TM AND TE MODES

Examining the components of the obtained expressions (1 – 4, 11, 12) for the fields of the TM and TE modes separately, we find that the transverse components have a divergence of the order of $\nu_m^{-2}$ at $\nu_m \to 0$ (without considering the features of the yet unknown function $\chi_m$) for arbitrary integer values $m > 0$. But due to the superposition (Eqn. 2) these singularities are mutually compensated for $m > 1$. Only when $m = 1$ a logarithmic divergence occurs at $\nu_1 \to 0$. Note, however, that no singularities appear in the longitudinal components.

Let us consider this issue in more detail and clarify the mechanism of the emergence of the singularities in the transverse components of the TM and TE modes and their mutual compensation during the summation. Near the singular point, the transverse electrical components of the TM and TE modes are represented for $m > 1$ as follows:

$$\left.\begin{array}{l}\mathcal{A}_m^{(J)} E_{m,TM_r}^{(J)} \\ \mathcal{A}_m^{(H)} E_{m,TM_r}^{(H)}\end{array}\right\}_{\nu_m \to 0} = \mp p_m^2 \alpha_m^{\pm} \nu_m^{-2}, \qquad \left.\begin{array}{l}\mathcal{B}_m^{(J)} E_{m,TE_r}^{J} \\ \mathcal{B}_m^{(H)} E_{m,TE_r}^{(H)}\end{array}\right\}_{\nu_m \to 0} = \pm k^2 \alpha_m^{\pm} \nu_m^{-2}$$

$$\left.\begin{array}{l}\mathcal{A}_m^{(J)} E_{m,TM_\varphi}^{(J)} \\ \mathcal{A}_m^{(H)} E_{m,TM_\varphi}^{(H)}\end{array}\right\}_{\nu_m \to 0} = -j \alpha_m^{\pm} p_m^2 \nu_m^{-2}, \qquad \left.\begin{array}{l}\mathcal{B}_m^{(J)} E_{m,TE_\varphi}^{J} \\ \mathcal{B}_m^{(I)} E_{m,TE_\varphi}^{(I)}\end{array}\right\}_{\nu_m \to 0} = j \alpha_m^{\pm} k^2 \nu_m^{-2}. \tag{13}$$

$$\alpha_m^{\pm} = m \chi_m \frac{q}{2\varepsilon_0} \frac{\omega_0}{\omega} \left(\frac{r}{a}\right)^{\pm m - 1}$$

As a result of the summation of the components of the TM and TE modes, a common factor $k^2 - p_m^2 = \nu_m^2$ arises, which compensates the singularities in the corresponding total multipole components. The expansion of the sum in the vicinity of $\nu_m = 0$ can be written in a form that does not contain a singularity:

$$\left.\begin{array}{l}\mathcal{E}_{m_r}^{(J)} \\ \mathcal{E}_{m_r}^{(H)}\end{array}\right\}_{\nu_m \to 0} = j \left.\begin{array}{l}\mathcal{E}_{m_\varphi}^{(J)} \\ \mathcal{E}_{m_\varphi}^{(H)}\end{array}\right\}_{\nu_m \to 0} = j \alpha_m^{\pm} + O[\nu_m]^0 . \tag{14}$$

The term $O[\nu_m]^0$ indicates the presence of additional terms in the expansions that do not contain the parameter $\nu_m$ (of order $[\nu_m]^0$).

Transverse magnetic TM and TE components in the vicinity of $\nu_m = 0$ have singularities equal in magnitude and opposite in sign:

$$\left.\begin{array}{l}\mathcal{A}_m^{(I)} H_{m,TM_r}^{(,I)} \\ \mathcal{A}_m^{(H)} H_{m,TM_r}^{(H)}\end{array}\right\}_{\nu_m \to 0} = \left.\begin{array}{l}-\mathcal{B}_m^{(I)} H_{m,TE_r}^{(,I)} \\ -\mathcal{B}_m^{(I)} H_{m,TE_r}^{(,I)}\end{array}\right\}_{\nu_m \to 0} = j \varepsilon_0 \omega p_m \alpha_m^{\pm} \nu_m^{-2}, \tag{15}$$

$$\left.\begin{array}{l}\mathcal{A}_m^{(I)} H_{m,TM_\varphi}^{(,I)} \\ \mathcal{A}_m^{(H)} H_{m,TM_\varphi}^{(H)}\end{array}\right\}_{\nu_m \to 0} = \left.\begin{array}{l}-\mathcal{B}_m^{(I)} H_{m,TE_\varphi}^{(J)} \\ -\mathcal{B}_m^{(J)} H_{m,TE_\varphi}^{(J)}\end{array}\right\}_{\nu_m \to 0} = \varepsilon_0 \omega p_m \alpha_m^{\pm} \nu_m^{-2},$$

For the transverse electric components, divergences of the order of $\nu_1^{-2}$ in the total field are eliminated due to the identity $k^2 - p_1^2 = \nu_1^2$.

When $m = 1$, an additional logarithmic singularity arises at $\nu_1 = 0$ in the distributions of the transverse electric and magnetic components of the TM and TE modes.

In particular, for the radial electrical components we find:

$$\left.\begin{array}{l}\mathcal{E}_{1_r}^{(I)} \\ \mathcal{E}_{1_r}^{(H)}\end{array}\right\}_{\nu_1 \to 0} = \pm \chi_1 \frac{q}{2\varepsilon_0} \frac{\omega_0}{\omega} \begin{Bmatrix} 1 \\ a^2/r^2 \end{Bmatrix} + \chi_1 \frac{a^2 q}{4\varepsilon_0} \frac{\omega_0}{\omega} (k^2 + p_1^2) \begin{Bmatrix} ln(a\nu_1/2) \\ ln(r\nu_1/2) \end{Bmatrix}. \tag{16}$$

For the azimuthal component $\mathcal{E}_{1_\varphi}^{(Z)} = j\mathcal{E}_{1_r}^{(Z)}$ with the positive sign of the first term holds.

The transverse magnetic components only contain the logarithmic term, the other term cancels due to the opposite sign and the equality of the amplitudes in the sum.

For the radial magnetic components:

$$\left.\begin{array}{l}\mathcal{H}_{1_r}^{(I)} \\ \mathcal{H}_{1_r}^{(H)}\end{array}\right\}_{\nu_1 \to 0} = -\chi_1 j \frac{qa^2}{2} p_1 \omega_0 \begin{Bmatrix} ln(a\nu_1/2) \\ ln(r\nu_1/2) \end{Bmatrix} \tag{17}$$

for azimuthal magnetic components $\mathcal{H}_{1_\varphi}^{(Z)} = j\mathcal{H}_{1_r}^{(Z)}$ holds.

Two conclusions can be drawn from the discussion above: 1) TM and TE modes are mutually coupled and cannot be generated separately. Mathematically, this manifests in the presence of a singularity of the order of $\nu_m^{-2}$ for TM and TE modes and its absence when both modes are generated simultaneously. 2) The logarithmic singularity that remains in the sum of the modes at $m = 1$ must be eliminated by an appropriate selection of the factor $\chi_m$.

## 4. DETERMINATION OF THE FACTOR $\chi_m$

The factor $\chi_m$ can in principle be determined by an appropriate multipole expansion of the three-dimensional delta-like charge and current density functions, but for a helical trajectory this is a challenging task.

Another possibility – which we will follow below – is to determine $\chi_m$ by comparison with one of the existing approximate solutions of the problem [1, 5], which is valid in the far-field and for large $\gamma$.

A suitable choice are the well-known relations obtained by Kincaid (see [5], Eqn. 27), describing the spectral radiation density of an extended helical undulator:

$$\tilde{I}(\omega) = \frac{Nq^2K^2\tilde{r}}{\varepsilon_0 c}\sum_{m=1}^{\infty}\left[J_m'^2(\tilde{x}_m) + \left(\frac{\tilde{\alpha}_m}{K} - \frac{m}{\tilde{x}_m}\right)^2 J_m^2(\tilde{x}_m)\right]\tilde{u}(\tilde{\alpha}_m^2). \quad (18)$$

Here $K$ is the undulator deflection parameter, $N$ denotes the number of periods of the undulator, $\tilde{u}(x)$ is the unit step function, and

$$\tilde{\alpha}_m^2 = m/\tilde{r} - 1 - K^2,\ \tilde{x}_m = 2K\tilde{r}\tilde{\alpha}_m,\ \tilde{r} = \omega/2\gamma^2\omega_0. \quad (19)$$

The relation, equivalent to Eqn. 18 can be obtained by using Eqns. 2, 11 and Eqn. 12, which are written in a cylindrical coordinate system.

The total radiation energy of the $m^{th}$ term of the expansion is expressed in the following integral form:

$$E_m = \lim_{L\to\infty}\frac{1}{L}\int_{-L/2}^{L/2}\int_{-\infty}^{\infty}\int_0^{2\pi}\int_{-\infty}^{\infty} I_m(\omega)dp_m r d\varphi \mathrm{v}_z dt dz = \int_{-\infty}^{\infty} J_m(\omega)d\omega \quad (20)$$

where

$$J_m(\omega) = \frac{4\pi^2 rN}{\omega_0}I_m(\omega) \quad (21)$$

is the spectral density of energy, radiated by $m^{th}$ multipole, and

$$I_m(\omega) = \varepsilon_0\vec{\mathcal{E}}_m^{(H)}\vec{\mathcal{E}}_m^{(H)^*} + \mu_0\vec{\mathcal{H}}_m^{(H)}\vec{\mathcal{H}}_m^{(H)^*}, \quad (22)$$

where $\varepsilon_0$ and $\mu_0$ are the electric and magnetic permeabilities of vacuum.

In Eqn. 20 the integration over $\mathrm{v}_z dt$ is performed over the actual length of the undulator $Nl$ ($l = 2\pi\mathrm{v}_z/\omega_0$ is the undulator period), and the integration over $p_m$ is replaced by an integration over $\omega$ by substituting $p_m = (\omega - m\omega_0)/\mathrm{v}_z$.

In the far zone, for $r \gg a$, the Hankel functions in Eqn. 22 can be represented by their asymptotic expression [13]:

$$\left[H_m^{(1)}(\nu_m r)\right]^2 = \left[H_m^{(1)'}(\nu_m r)\right]^2 \approx 2/\pi\nu_m r. \quad (23)$$

Higher terms of order $r^{-1}$ and higher can be discarded. As a result, the new version of the expression for the spectral energy distribution density takes the form:

$$J(\omega) = \frac{q^2N}{2\varepsilon_0 c}\tilde{\omega}\sum_{m=1}^{\infty} Q_m\left\{\frac{\tilde{D}_m^2}{S_m}J_m^2(\tilde{y}_m) + \beta_\varphi^2 J_m'^2(\tilde{y}_m)\right\}\tilde{u}(s_m) \quad (24)$$

where

$$\begin{gathered}\tilde{\omega} = \omega/\omega_0,\\ Q_m = 4\pi^3\chi_m^2 a^2\omega/c\nu_m,\\ \tilde{D}_m = m - \tilde{\omega}/\gamma_z^2,\\ S_m = \tilde{\omega}\tilde{D}_m - m(m-\tilde{\omega}) = (1-\beta_z^2)s_m,\\ s_m = (\tilde{\omega} - m(1+\beta_z)^{-1})(m(1-\beta_z)^{-1} - \tilde{\omega})\\ \tilde{y}_m = \frac{\beta_\varphi}{\beta_z}\sqrt{S_m},\\ \gamma_z = (1-\beta_z^2)^{-1/2}\\ \beta_\varphi = \sqrt{\beta_\gamma^2 - \beta_z^2},\ \ \beta_z = \mathrm{v}_z/c,\ \ \beta_\gamma = \sqrt{1-\gamma^{-2}},\end{gathered} \quad (25)$$

The parameter $Q_m$ in Eqn. 24 is artificially assigned: it will have to remain unchanged during the further transformations.

To establish a correspondence of Eqn. 24 and Eqn. 18 approximate formulas for the longitudinal and transverse components of the particle velocity can be used. Expressed through the undulator deflection parameter $K$ [5] they read as:

$$\mathrm{v}_z = c\sqrt{(1-\gamma^{-2})(1-K^2\gamma^{-2})},\qquad \beta_\varphi = \frac{K}{\gamma}\frac{\mathrm{v}_z}{c},\ \ a = \frac{K\mathrm{v}_z}{\gamma\omega_0}. \quad (26)$$

Substituting the first of Eqns. 26 into the coefficient $\frac{\tilde{D}_m^2}{S_m}$ (Eqn. 24), we find:

$$\frac{\tilde{D}_m^2}{S_m} = \frac{\left(\gamma^4 m + K^2\tilde{\omega} - \gamma^2(1+K^2)\tilde{\omega}\right)^2}{\gamma^8 m(2\tilde{\omega}-m) + \gamma^4 K^2\tilde{\omega}^2 - \gamma^6(1+K^2)\tilde{\omega}^2}. \tag{27}$$

Assuming $\gamma^2 \gg 1$, the term $K^2\tilde{\omega}$ is small in comparison to $K^2\tilde{\omega}\gamma^2$ in the numerator of Eqn. 27, and can be neglected. Equally the term $\gamma^4 K^2\tilde{\omega}^2$ can be neglected in comparison to $\gamma^6 K^2\tilde{\omega}^2$ in the denominator. In addition, by assuming $2\tilde{\omega} \gg m$, $m$ can be ignored in the denominator. As result we find:

$$\frac{\tilde{D}_m^2}{S_m} \approx \frac{1}{\gamma^2}\frac{\left(\gamma^2 m - (1+K^2)\tilde{\omega}\right)^2}{2\tilde{\omega}\gamma^2 m - (1+K^2)\tilde{\omega}^2} = \frac{K^2}{\gamma^2}\left(\frac{\tilde{\alpha}_m}{K} - \frac{m}{\tilde{x}_m}\right)^2 \tag{28}$$

The coefficient $\beta_\varphi^2$ in Eqn. 24 turns into $\frac{K^2}{\gamma^2}$, if we assume $\mathrm{v}_z = c$, which is possible at relativistic energies.

Finally, argument $\tilde{y}_m$ in Eqn. 24 is represented by:

$$\tilde{y}_m^2 = -\frac{\tilde{\omega}^2 K^2\left(\gamma^2 + (\gamma^2-1)K^2\right)}{\gamma^6} + \frac{2K^2 m\tilde{\omega}}{\gamma^2} - \frac{K^2 m^2}{\gamma^2} \tag{29}$$

where the undulator parameter $K$ has been used.

For $2\tilde{\omega} \gg m$ the third term in Eqn. 29 can be neglected in comparison to the second term, and for $\gamma^2 \gg 1$, $\gamma^2 - 1 = \gamma^2$ in the numerator of the first term. As a result, we obtain an approximate expression for $\tilde{y}_m$ that coincides with Kincaid's notation (cf. Eqn. 19):

$$\tilde{y}_m \approx \frac{\omega K}{\gamma^2\omega_0}\sqrt{\frac{2\gamma^2 m\omega_0}{\omega} - 1 - K^2} = \tilde{x}_m \tag{30}$$

Substituting the obtained approximate expressions for the coefficients and the argument into Eqn. 24 transforms it into an expression, which is identical to Kincaid relation Eqn. 18, with the additional condition that $Q_m = 1$.

The desired value of the parameter $\chi_m$ follows hence from $Q_m = 1$ as:

$$\chi_m = \frac{1}{2a\pi^{3/2}}\left(\frac{c\nu_m}{\omega}\right)^{1/2} \tag{31}$$

For $\omega \gg m\omega_0$ and (or) $\gamma \gg 1$ it turns into a complex constant independent of $m$, whereas for

$$m(1+\beta_z)^{-1} < \omega/\omega_0 < m(1-\beta_z)^{-1} \tag{32}$$

($\nu_m > 0$) it is a positive real function versus $\omega$, which eliminates the logarithmic singularity of the transverse field components at $\nu_1 = 0$ for $m = 1$.

It should be emphasized that the simplifying transformations (Eqns. 26 – 30) in the transition from Eqn. 24 to Eqn. 18 did not affect the expression for $Q_m$ (cf. Eqn. 25). Thus, Eqn. 24 retained its original form, including the initial components of the exact solution, which does not contain the approximations made in Eqns. 26.

Taking into account Eqn. 31, the amplitudes Eqn. 11 can now be reduced to their final form:

$$\begin{matrix}\mathcal{A}_m^{(J)} \\ \mathcal{A}_m^{(H)}\end{matrix} = \frac{q}{4\varepsilon_0 V\omega}\left(\frac{c\nu_m}{\pi\omega}\right)^{1/2} f_m \begin{Bmatrix} H_m^{(1)}(\tilde{y}_m) \\ J_m(\tilde{y}_m)\end{Bmatrix}, \quad \begin{matrix}\mathcal{B}_m^{(J)} \\ \mathcal{B}_m^{(H)}\end{matrix} = j\frac{aq\omega_0}{4\varepsilon_0 c}\left(\frac{c\nu_m}{\pi\omega}\right)^{1/2} \nu_m \begin{Bmatrix} H_m^{(1)\prime}(\tilde{y}_m) \\ J_m'(\tilde{y}_m)\end{Bmatrix} \tag{33}$$

For arbitrary particle energy $\gamma$ and longitudinal velocity $\mathrm{v}_z$, expressions (1 – 4) in combination with Eqn. 33 can be used to construct the distributions of the field components and other characteristics of the radiation in an arbitrary region of space.

## 5. COMPARISON OF EQUATIONS 18 AND 24

Eqn. 24 with $Q_m = 1$ describes the spectral density of the emitted energy as a function of the normalized frequency $\tilde{\omega} = \omega/\omega_0$ with a parametric dependence on the energy $\gamma$ and the longitudinal velocity of the particle $\mathrm{v}_z$.

Eqn. 18 has one remarkable property: it allows the spectra to depend on the generalized frequency parameter $\tilde{r} = \omega/2\gamma^2\omega_0$, leaving only the parameter $K$ free. As shown with Eqns. 28 – 30, this is achieved in the approximation of relativistic energies ($\gamma \gg 1$) and far from frequencies that are multiples of the rotation frequency ($2\omega \gg m\omega_0$).

Substituting the approximate expression for the longitudinal velocity (Eqn. 26) into Eqn. 24, turns it into a refined version of Eqn. 18: in this case, the requirement for $\gamma$ is relaxed and the constraint $2\omega \gg m\omega_0$ is completely eliminated. This option retains the parametric dependence on $K$ in Eqn. 24 with the formal introduction of a functional dependence on $\tilde{r}$, as in Eqn. 18, while acquiring an additional parametric dependence on γ.

The latter makes it possible to superimpose the spectra calculated with Eqn. 18, which does not change with $\gamma$, with curves calculated with Eqn. 24 with $Q_m = 1$, for different energies and to trace the evolution of their shape as the Lorentz factor $\gamma$ changes.

Figures 2 and 3 display distributions of the spectral density of the radiation energy of a particle moving along a helical trajectory. The red dotted curve is calculated using Eqn. 18, the black solid curves, calculated for three different values of γ, correspond to Eqn. 24. While the top left panel shows the full spectrum, the other panels display specific parts of the spectrum as magnified sections. The intensity $\tilde{I}$ is normalized to $Nq^2/\varepsilon_0 c$.

The comparison is carried out for an undulator parameter above 1, ($K = 2$, Fig.2), and a small undulator parameter below 1, ($K = 0.25$, Fig.3); the Lorentz factor γ is varied between 3 and 10, i.e. 1.5 to 5 MeV electron beam energy, while Eqn. 18 is valid only for $\gamma \gg 1$.

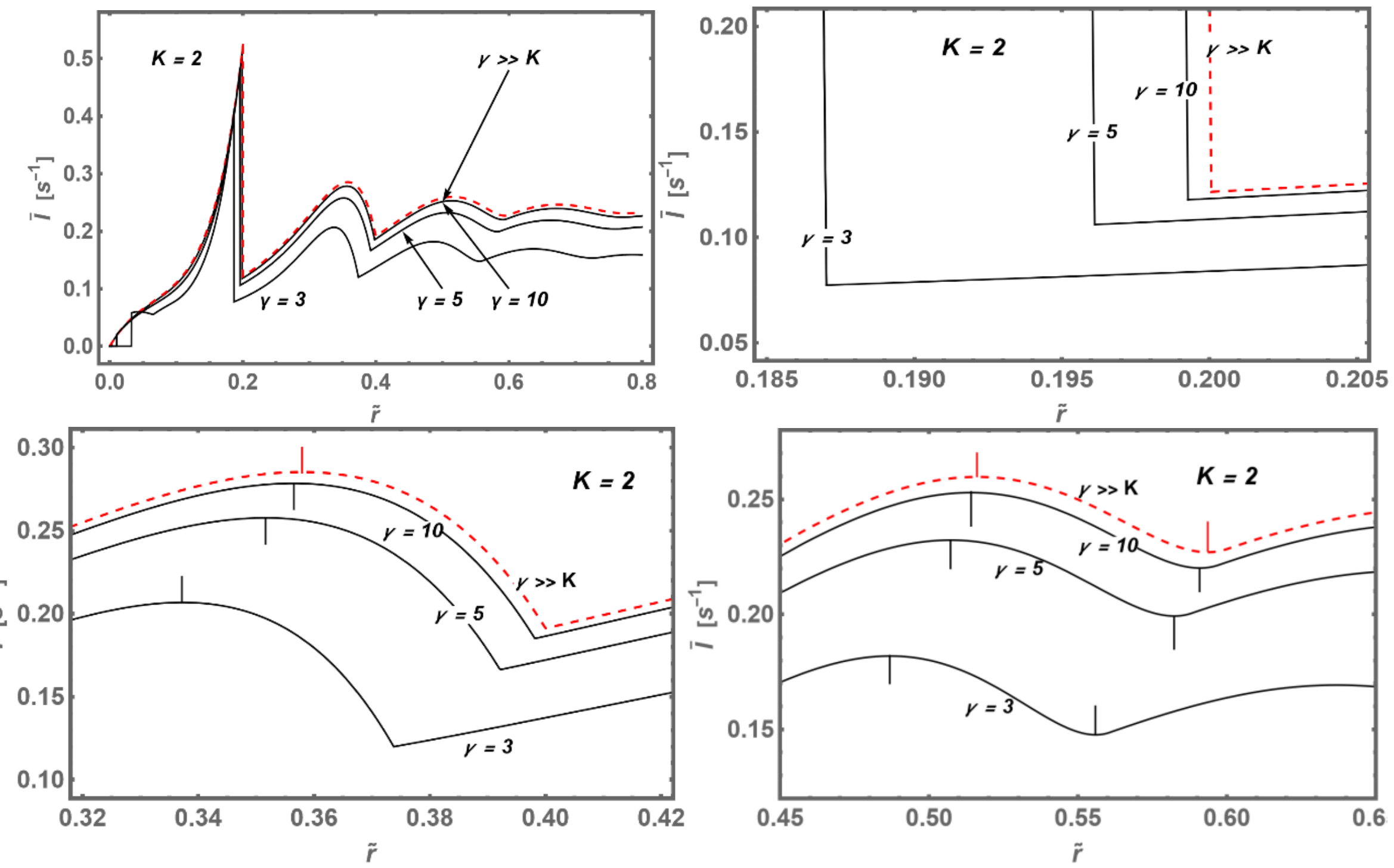

Figure 2: Comparison of the normalized spectral density of the radiation energy of a particle moving along a helical trajectory. The red dotted line is the approximate solution (Eqn. 18), while the black lines show the result of the exact solution (Eqn. 24) for three values of the Lorentz factor. The undulator parameter is $K = 2$, the top left panel shows the complete spectrum, the other panels show magnified sections. Local maxima are indicated by vertical line segments in the lower row panels.

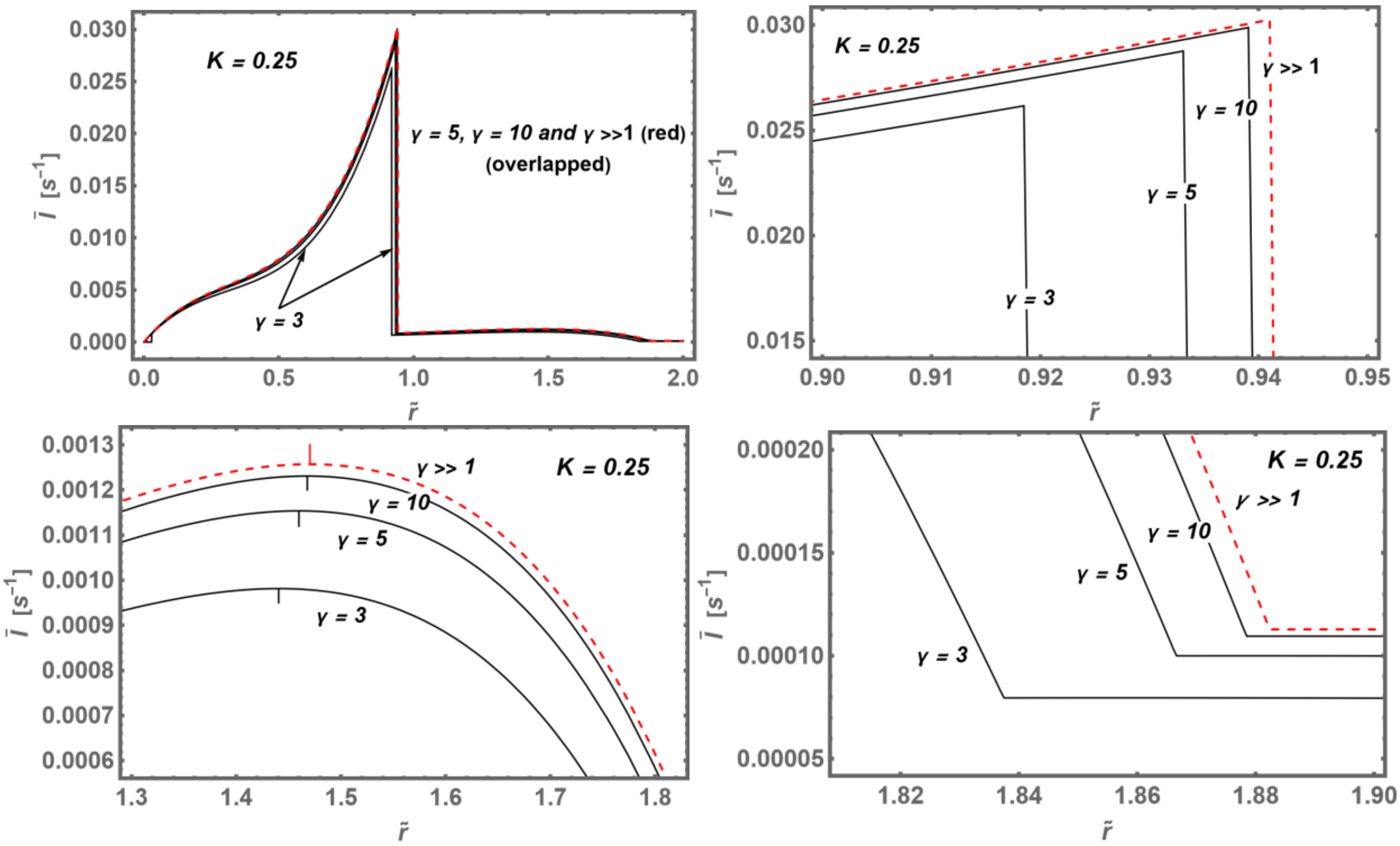

Figure 3: Comparison of the normalized spectral density of the radiation energy of a particle moving along a helical trajectory. The red dotted line is the approximate solution (Eqn. 18), while the black lines show the result of the exact solution (Eqn. 24) for three values of the Lorentz factor. The undulator parameter is $K = 0.25$, the top left panel shows the complete spectrum, the other panels show magnified sections. Local maxima are indicated by vertical line segments in the lower row panels.

The comparison demonstrates a general decrease of the levels of the distributions when Eqn. 24 is employed, rather than the approximate solution Eqn. 18. It is especially noticeable at low energies, i.e., small values of $\gamma$, and comparatively large values of $K$. Also, the position of the critical frequencies, visible as characteristic sharp kinks in the spectrum, is displaced toward lower frequencies with decreasing Lorentz factor (top and bottom right). Equivalently the position of the local maximum of the second harmonics shifts downward (bottom left). The reason for these trends is that in Eqn. 24 the exact parameter $\tilde{y}_m$ is used as argument in the Bessel functions, while Eqn. 18 is based on approximate the values $\tilde{x}_m$.

It should be recalled that Eqn. 24 can be used without specification of the undulator parameter $K$. To unambiguously construct spectra, it is sufficient to specify two parameters: the total energy of the particle $\gamma$ and the longitudinal component of its velocity $\beta_z = v_z/c$. In this case, it is not necessary to impose the condition $\gamma \gg 1$ and in this regard, the presentation of the spectral distributions for $\gamma = 3$ and $\gamma = 5$, which stretches the condition $\gamma \gg 1$ (especially for $K = 2$), does not seem to be problematic.

In addition to the observed trends a reduction of the allowed frequency band for a given mode is found with decreasing in the longitudinal velocity component when Eqn. 24 is applied, cf. Eqn. 32.

For Eqn. 18, the allowed frequency band of the distributions is determined by the inequality

$$0 < \tilde{\omega} < \frac{2\gamma^2 m}{1+K^2}. \quad (34)$$

The allowed region of the spectrum is thus limited only to the high frequencies side, and, just as in Eqn. 24, the width of the allowed band increases with increasing particle energy and the number of modes $m$.

However, the absence of a restriction on the low frequency side for Eqn. 18 leads to a sequential superposition of the spectra of individual modes, resulting in a consistent increase in the level of the total distribution.

In case of applying the Eqn. 24 the allowed frequency regions of the spectrum are determined from condition Eqn. 32, which implies the narrowing of the frequency bands on the high and the low frequency side as the longitudinal propagation speed decreases. Limitation of the allowed frequency range from low frequencies can lead to jumps in the distribution of spectral radiation energy density (see, for example, the curves for $\gamma = 3$ in Figs. 2, 3, top, left). Jumps are characteristic of the boundary frequencies of the allowed interval of the fundamental mode $m = 1$: only this mode gives finite values for the distribution (18) at $\tilde{\alpha}_1 = 0$ and the distribution (24) at $\nu_1 = 0$. For Eqn. 18, this is the upper limit of the interval $\tilde{\omega} = 2\gamma^2(1+K^2)^{-1}$, whereas for Eqn. 24, jumps occur at both the upper $\tilde{\omega} = (1-\beta_z)^{-1}$ and lower $\tilde{\omega} = (1+\beta_z)^{-1}$ boundary frequencies. Additional low-frequency jumps, marking the minimum initial frequency of undulator radiation, appear in Figures 2 and 3.

## 6. POINTING VECTOR AND DOPPLER EFFECT

The direction of radiation propagation is determined by the orientation of the Poynting vector [14]. Its modal far-field components versus frequency in a unit vectors of cylindrical coordinate system are given below:

$$\vec{S}_m = \frac{1}{2}\left\{\vec{\mathcal{E}}_m^{(J)} \times \vec{\mathcal{H}}_m^{(J)^*} + \vec{\mathcal{E}}_m^{(J)^*} \times \vec{\mathcal{H}}_m^{(J)}\right\} = S_{m,r}\vec{e}_r + S_{m,z}\vec{e}_z$$

$$S_{m,r} = q^2 U_m,\ S_{m,z} = q^2 U_m \frac{\omega - m\omega_0}{v_m v_z} \qquad (35)$$

$$|\vec{S}_m| = \frac{\omega}{c v_m} U_m$$

$$U_m = \frac{c^2 f_m^2 J_m^2(a v_m) + J_m'^2(a v_m) a^2 v_m^2 v_z^2 \omega_0^2 \omega^2}{8\pi^2 r c \varepsilon_0 v_m v_z^2 \omega^2}$$

When passing through the frequency $\omega = m\omega_0$, the longitudinal component of the Poynting vector changes sign: at $\omega > m\omega_0$ it is positive, i.e., we are dealing with forward radiation, otherwise $(\omega < m\omega_0)$ there is backward radiation. Thus, in accordance with the Doppler effect, the high-frequency part of the radiation is directed forward, and the low-frequency part is directed backward. At $\omega = m\omega_0$, the longitudinal component vanishes and radiation occurs only in the radial direction. At $v_m = 0$ (at the boundary points of the allowed frequency interval) the radial component of the Poynting vector vanishes for all positive integer values of $m$, and the longitudinal component is not equal to zero only at $m = 1$. Thus, only the fundamental dipole mode $m = 1$ is responsible for coaxial radiation (directed both forward and backward).

The ratio of the radial and longitudinal components of the Poynting vector gives the tangent of the angle between the direction of radiation propagation and the longitudinal axis $z$:

$$tg\vartheta = \frac{S_{m,\varphi}}{S_{m,z}} = \frac{v_m v_z}{\omega - m\omega_0} = \frac{\sqrt{(\tilde{\omega} - m(1+\beta_z)^{-1})(m(1-\beta_z)^{-1} - \tilde{\omega})}}{\gamma_z(\tilde{\omega} - m)} \qquad (36)$$

For a fixed angle $\theta$, Eqn. 36 determines the frequency $\omega_m$ corresponding to a given mode, recorded by an observer located in a given direction

$$\omega_m = \frac{m\omega_0}{1 - \beta_z cos\theta} \qquad (37)$$

and at a fixed frequency $\omega$ it determines the directions $\theta_m$ along which the observer recording the given frequency must be located:

$$cos\theta_m = \frac{1}{\beta_z}\left(1 - \frac{m\omega_0}{\omega}\right) \qquad (38)$$

Eqns. 37 and 38 together represent explicit expressions for the Doppler effect as applied to the modal components of the radiation field of a helical undulator. At $\beta_z = 0$, i.e., in a rest frame relative to the particle (motion along a closed circle), the frequency does not depend on the observation angle and for a given mode acquires a fixed value $\omega_m = m\omega_0$.

Angles $0 \le \theta_m(\theta) < \pi/2$ correspond to radiation directed forward (going ahead of the particle), and at angles $\pi/2 < \theta_m \le \pi$ the direction of radiation is opposite to the movement of the particle. At angles $\theta_m(\theta) = 0$ and $\theta_m(\theta) = \pi$, the radiation is directed parallel to the axis $z$ at the boundary frequencies of the permitted frequency intervals (32).

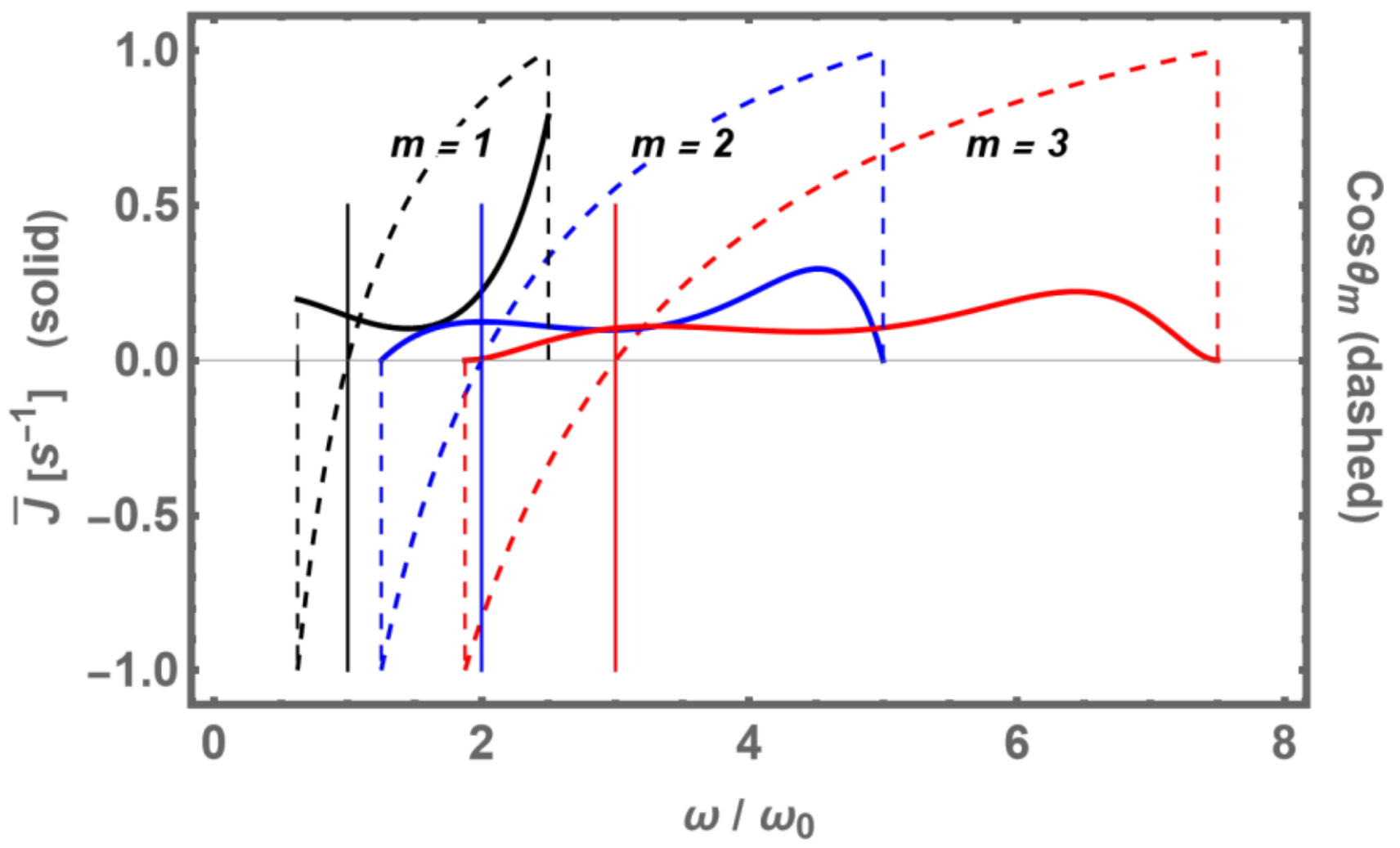

Figure 4. Distributions of spectral densities of radiation energy of the first three modes of a helical undulator; $\gamma = 10, \beta_z = 0.6$

In the general case, each frequency (within the permitted frequency range) corresponds to a certain direction in space, determined by the relation (38). Figure 4 shows the spectral distributions of the radiation energy density of the first three modes (the first three terms of the sum in Eqn. 24). In parallel, curves describing the function $cos\theta_m$ (38) depending on the normalized frequency $\tilde{\omega}$ are superimposed on the graphs of spectral distributions. The vertical dotted lines establish correspondences between the values of the spectral distributions at the extreme points of the permitted regions and the values of the angles $\theta_m$ characterizing the direction of radiation propagation. The segments of solid vertical lines correspond to integer values of normalized frequencies $\tilde{\omega}$. They intersect the cosine curves at points $cos\theta_m = 0$ and divide the distribution curves into sections responsible for forward ($m < \tilde{\omega} < m(1-\beta_z)^{-1}$) and backward ($m(1+\beta_z)^{-1} < \tilde{\omega} < m$) radiation. The graphical constructions in Fig. 4 confirm the previously made statements about the formation of coaxial radiation by the extreme frequencies of the fundamental mode $m = 1$.

## 7. SPECTRUM DISCRETIZATION

A comparison of the modal distributions versus normalized frequency $\tilde{\omega}$, obtained using Eqn. 24 (Fig. 4) with similar distributions constructed using Eqn. 18 is presented in Fig. 5. When constructing the latter, the same energy ($\gamma = 10$) is used, and the parameter К is determined from the agreement of $\beta_z = 0.6$ with the approximate expression for the longitudinal velocity $v_z$ (Eqn. 24). It turns out to be equal to $К \approx 7.977$.

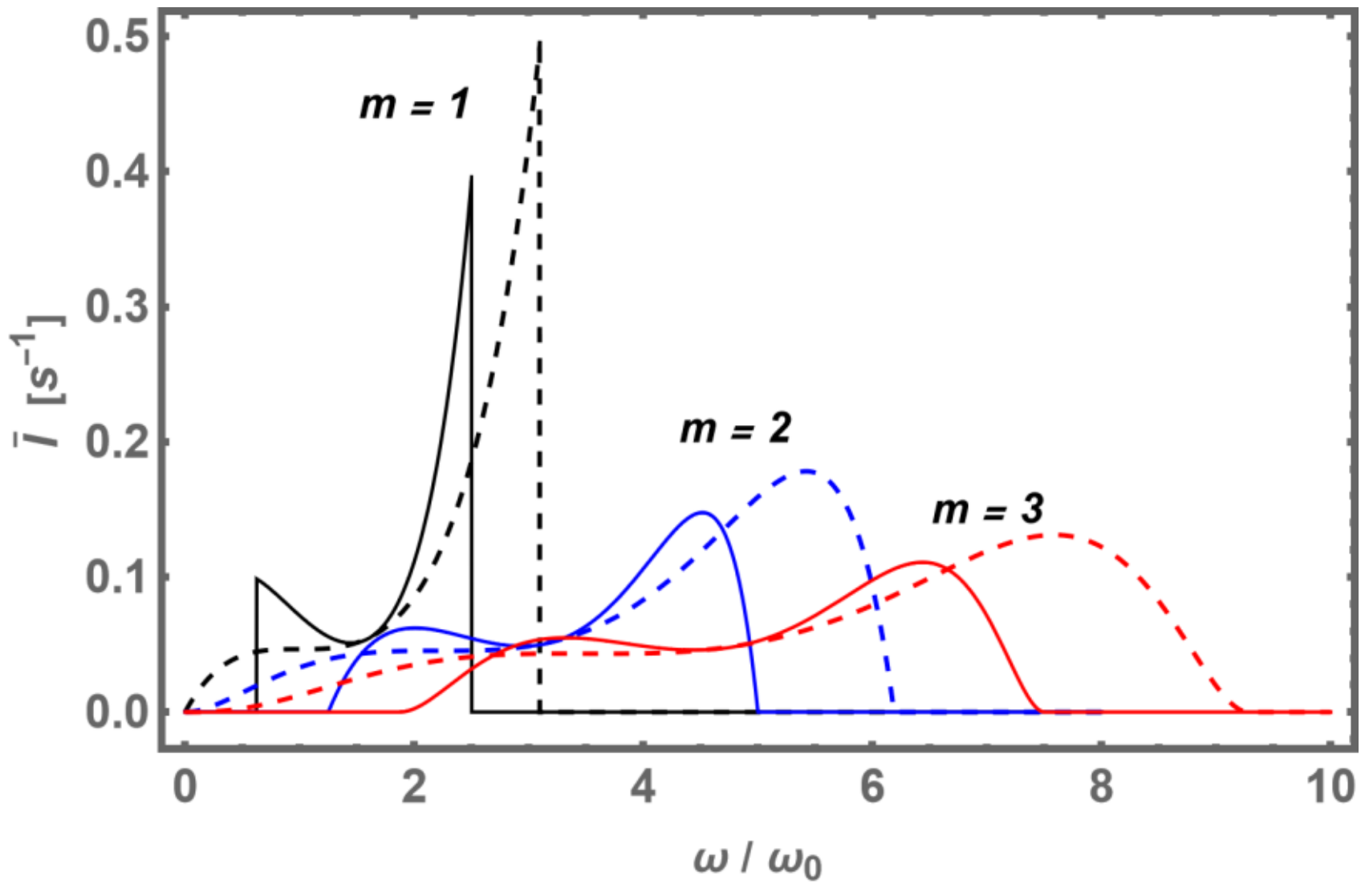

Figure 5. Comparison of modal distributions calculated using Eqns. 24 (solid, $\gamma = 10, \beta_z = 0.6$) and 18 (dashed, $\gamma = 10, К \approx 7.977$).

The significant differences between the exact solution (solid curves) and the solution constructed using the approximate Eqn. 18 are twofold. The zero lower limit for the allowed regions for modes of arbitrary order follows from condition of Eqn. 34, which is formed from requirement $2\tilde{\omega} \gg m$ when moving from the exact expression for $\tilde{y}_m$ (25), used in Eqn. 24 to the approximate form $\tilde{x}_m$ (19), used in (18). Therefore, all modes in approximation (18) originate from zero frequency regardless of the particle energy $\gamma$ and the value of the deviation parameter K.

Another noticeable difference is the broadening of the spectral bands, accompanied by a shift of local maxima towards high frequencies. This occurs due to too large values of the deflection parameter $K$ (for such values of $K$, Eqn. 18 is not provided) and a comparatively large $K/\gamma$ ratio (the requirement $K/\gamma \ll 1$ is violated). The expediency of such comparisons (Fig. 5) is justified by the possibility of visually determining the limits of applicability of approximation (18).

From Figures 4 and 5 it can be seen that the allowed frequency regions of the spectra of adjacent modes, determined by Eqn. 32, partially overlap each other but at sufficiently low longitudinal velocities, the allowed regions of the spectra of neighbouring modes cease to intersect and the radiation spectrum for several lower modes it becomes fragmented or quasi-discrete. An example of such a fragmented distribution is shown in Figure 6. The special role of the fundamental mode at $m = 1$ is retained even for arbitrarily small $\beta_z$ (see Fig. 7).

Figure 7 allows us to trace the process of successive narrowing of the spectra of individual modes with a gradual decrease in the longitudinal velocity of particle propagation $\beta_z$ (from $\beta_z = 0.05$ to $\beta_z = 0.01$). Along with the spectral distributions for the modes with numbers $m = 1,2$ and 10 for three different values of $\beta_z = 0.01, 0.025$ and $0.05$, the distributions (nominated on $m$) of the function $cos\theta_m$ are also given here, indicating the spatial directivity of the radiation corresponding to each frequency for all three values of $\beta_z$, distinguishing in this case (at $cos\theta_m = 0$ or $\theta_m = \pi/2$ and $\tilde{\omega} = m$) high frequencies ($\tilde{\omega} > m$), at which the radiation is directed forward ($0 \leq \theta_m < \pi/2$) and low frequencies ($\tilde{\omega} < m$) with radiation directed backward ($\pi/2 < \theta_m \leq \pi$). The unique property of the fundamental mode, noted by us at comparatively large values of $\beta_z$ (see Fig. 4), as follows from Figure 7, is preserved at arbitrarily small values of $\beta_z$: the maximum value of the frequency spectrum of the fundamental mode for all values of $\beta_z$ given in Figure 7 (top, left) still falls at the upper cut-off frequency $\tilde{\omega} = (1 - \beta_z)^{-1}$ and the spatial orientation of the radiation at this frequency corresponds to the direction parallel to the z axis ($\theta = 0$). The remaining modes do not emit radiation in this direction (see Fig. 7 for $m = 2$ and $m = 10$).

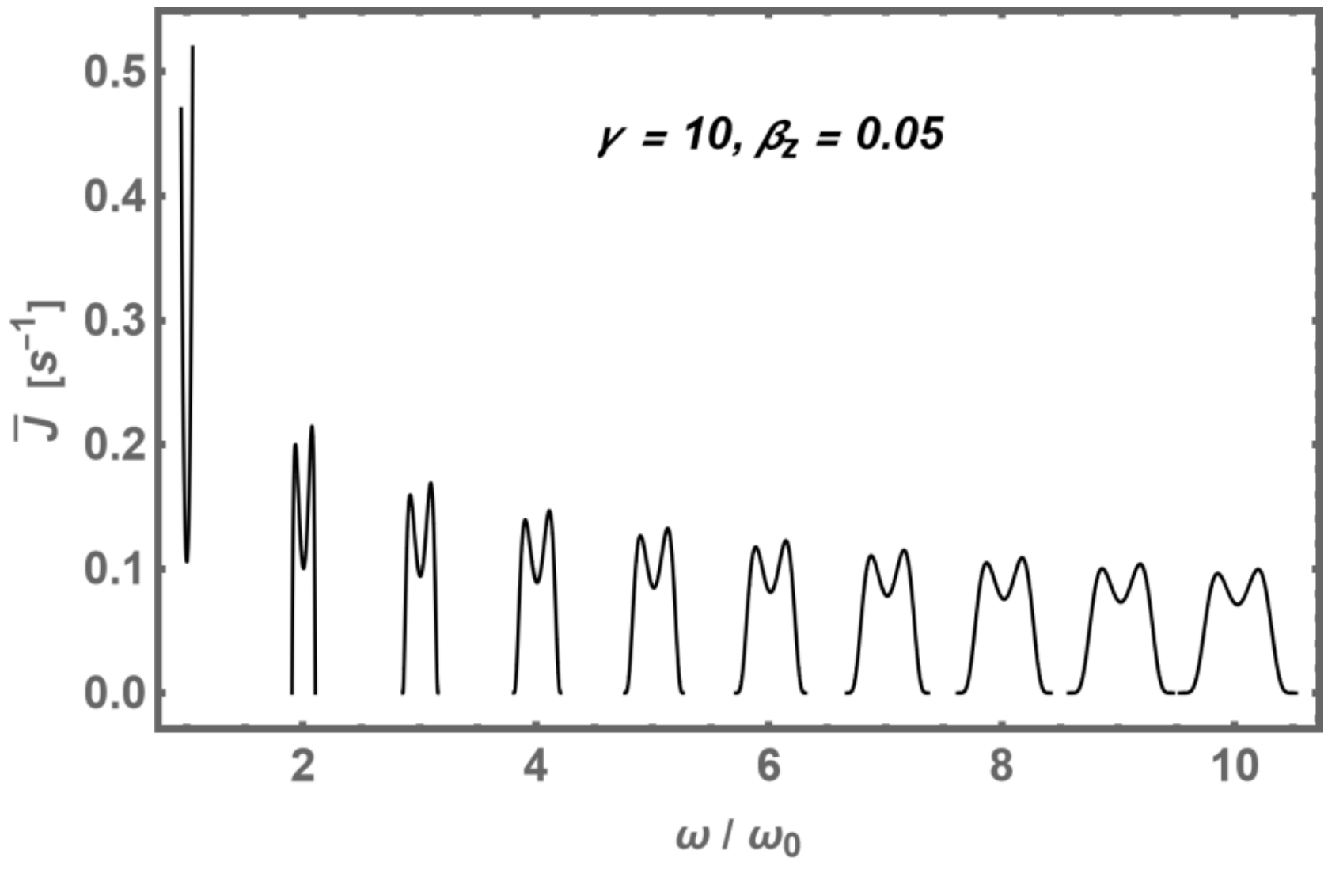

Figure 6. An example of spectrum discretization for small value of $\beta_z$.

Discretization of the radiation spectra by decreasing $\beta_z$ leads to the equalization of the low-frequency and high-frequency branches of the modal distributions (including the distribution of the fundamental mode), which leads to an increase in the rear radiation (almost 50% of the radiation goes back) and to some weakening of the amplitude of the fundamental mode responsible for radiation directly forward. An advantage of discretization is the narrowing of the modal spectral bands (including the fundamental mode), which may be will allow obtaining radiation of a high degree of monochromaticity in a narrow angular sector, oriented directly forward. In any case, there is a fundamental possibility of achieving this.

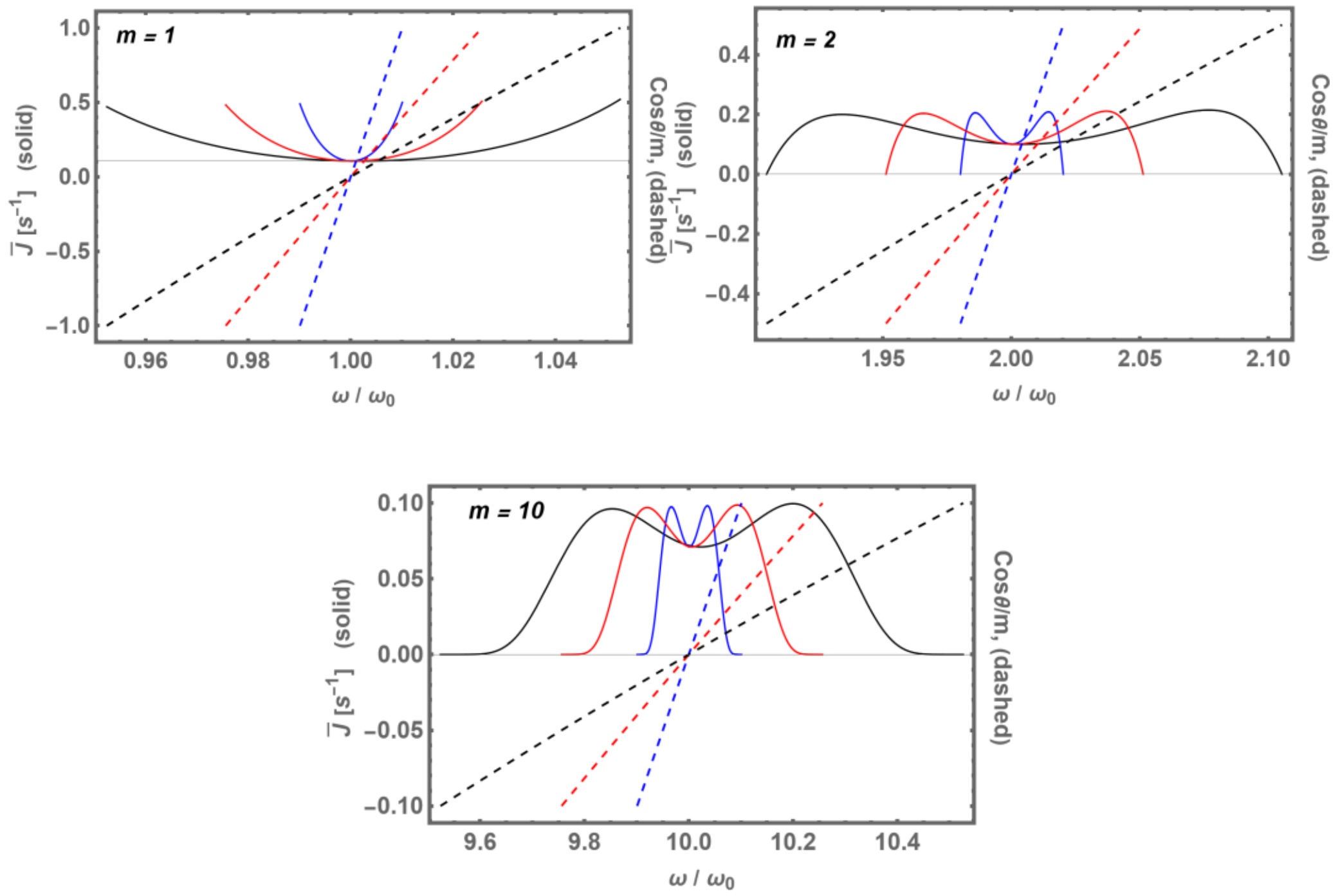

Figure 7. Distribution of the modal spectral density of the radiation energy of a particle moving along a spiral trajectory at low longitudinal velocities: $\beta_z = 10^{-2}$ (blue), $2.5 \cdot 10^{-2}$ (red), $5 \cdot 10^{-2}$ (black); first (top, left), second (top, right) and tenth (bottom) harmonics; $\gamma = 10$.

## 8. ANGULAR PROPERTIES OF RADIATION

Is it possible to make a direct transition in Eqn. 24 to the special case of particle motion along a closed circle ($\beta_z = 0$)? Direct substitution of $\beta_z = 0$ into Eqn. 36 yields an imaginary value for $tg\theta$, which indicates the impossibility of directly performing such a substitution. Such a transition means a transformation of the particle's trajectory: an infinite helical trajectory (which it is for arbitrarily small but not equal to zero $\beta_z$) at $\beta_z = 0$ turns into a trajectory of finite length in the form of a fixed closed circle.

To carry out such a transformation, we will use the Eqn. (37) for the Doppler effect, which relates the frequency $\omega$ in the laboratory system with the modal frequency $m\omega_0$ in the rest system of the particle's orbit at $\beta_z = 0$. Substituting expression (37) into the m-th term of the sum (24) (at $Q_m = 1$), we obtain the dependence of the m-th term of the multipole expansion of the spectral density of the radiation energy of a particle on the observation angle of the laboratory (stationary) frame of reference:

$$J_m(\beta_z, \theta) = \frac{q^2 N}{2\varepsilon_0 c} S_m(\beta_z, \theta) \tag{39}$$

$$S_m(\beta_z, \theta) = \frac{m}{1 - \beta_z cos\theta} \left\{ (ctg\theta - \beta_z cosec\theta)^2 J_m^2(\tilde{y}_m) + \beta_\varphi^2 J_m'^2(\tilde{y}_m) \right\}$$

Here now

$$\tilde{y}_m = \frac{\beta_\varphi m sin\theta}{1 - \beta_z cos\theta} \tag{40}$$

The transition to a frame of reference associated with a rotating orbit, i.e., to the rotation of a particle along a stationary closed circle, is accomplished by a simple substitution in Eqn. 39 and 40 $\beta_z = 0$ with $N = 1$ (spectral density of energy emitted per revolution):

$$J_m(0, \theta) = \frac{q^2 m}{2\varepsilon_0 c} \left\{ ctg^2\theta J_m^2(\beta_\gamma m sin\theta) + \beta_\varphi^2 J_m'^2(\beta_\gamma m sin\theta) \right\} \tag{41}$$

To determine the total spectral energy density of the m-th harmonic radiation, expressions (39) and (41) should be integrated over angle $\theta$. For helical motion of a particle, we have from (39):

$$J_m(\beta_z) = \frac{q^2 N}{2\varepsilon_0 c} S_m(\beta_z) \tag{42}$$

with

$$S_m(\beta_z) = m \int_0^{\pi} \frac{(ctg\theta - \beta_z cosec\theta)^2 J_m^2(\tilde{y}_m) + \beta_\varphi^2 J_m'^2(\tilde{y}_m)}{1 - \beta_z cos\theta} sin\theta d\theta \tag{43}$$

The limiting case of Eqn. 42 corresponds to the circular motion:

$$J_{,m}^0 = \frac{q^2}{2\varepsilon_0 c} S_m(0) \tag{44}$$

with

$$S_m^0 = S_m(0) = m \int_0^{\pi} \{ctg^2\theta J_m^2(\beta_\gamma m sin\theta) + \beta_\varphi^2 J_m'^2(\beta_\gamma m sin\theta)\} sin\theta d\theta \tag{45}$$

Figure 8 demonstrates the distributions of the spectral densities of the particle radiation energy per mode dependence on the value of the particle velocity longitudinal component $\beta_z$. Each curve corresponds to a specific value of $\beta_z$. The results of calculations for two different values of particle energy ($\gamma = 5$ and $\gamma = 10$) are presented, which allows us to draw some conclusions about the nature of the dependence of the considered indicator (function $S_m(\beta_z)$) on energy. Increasing energy contributes to a slower decrease in this indicator versus mode number. As can be seen from the comparison of the graphs, the level of the function $S_m(\beta_z)$ for a given m for $\gamma = 5$ approximately corresponds to the level of the mode with the number $4m$ for $\gamma = 10$. It is also obvious that the level of the function $S_m(\beta_z)$ decreases with increasing $\beta_z$. The maximum level is reached at $\beta_z = 0$. Directly adjacent to it from below is a curve corresponding to small value of $\beta_z = 0.2\beta_\gamma$, which indicates the presence of a direct transition from the case of motion along a closed circle ($\beta_z = 0$, a closed trajectory of finite length) to motion along a spiral ($\beta_z > 0$, an open trajectory of infinite length). It also follows from Figure 8 that the predominant contribution to the modal density of radiated energy comes from the fundamental mode ($m = 1$), and the smaller $\beta_z$, the greater the value of this contribution.

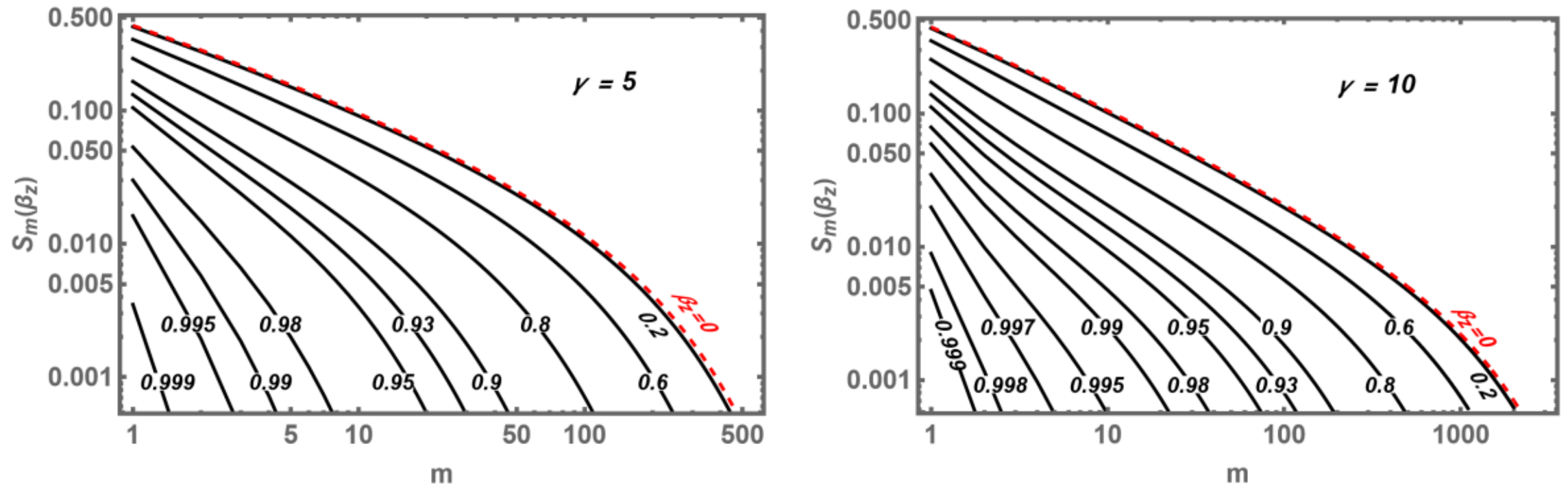

Figure 8. Modal distributions of radiated energy densities normalized to $q^2/2\varepsilon_0 c$ for different values of $\beta_z/\beta_\gamma$ (marked on the graphs by decimal fractions) for $\gamma = 5$ (left) and $\gamma = 10$ (right).

The characteristics shown in Fig. 8 record the total modal spectral energy density integrated in all directions. Polar diagrams constructed using Eqn. u 39 and 41, displaying the dependence of the spectral density of radiation energy on the direction in space, are shown in Figure 9.

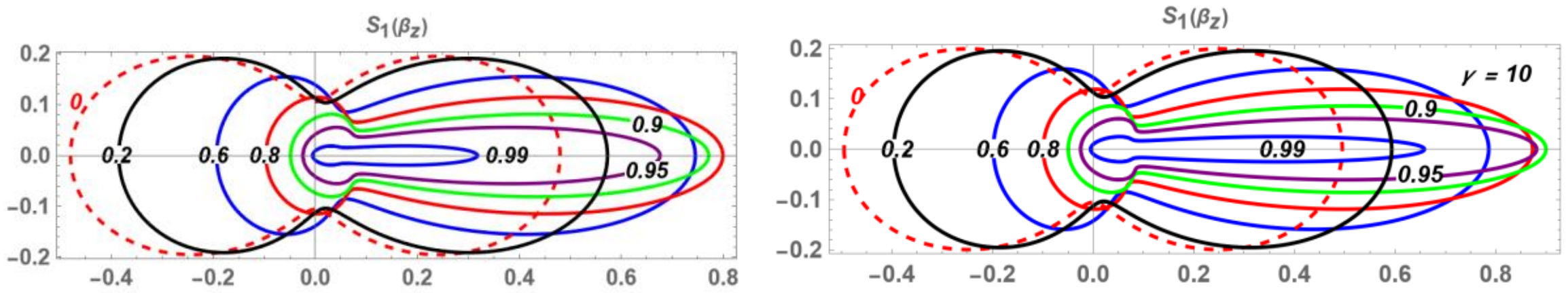

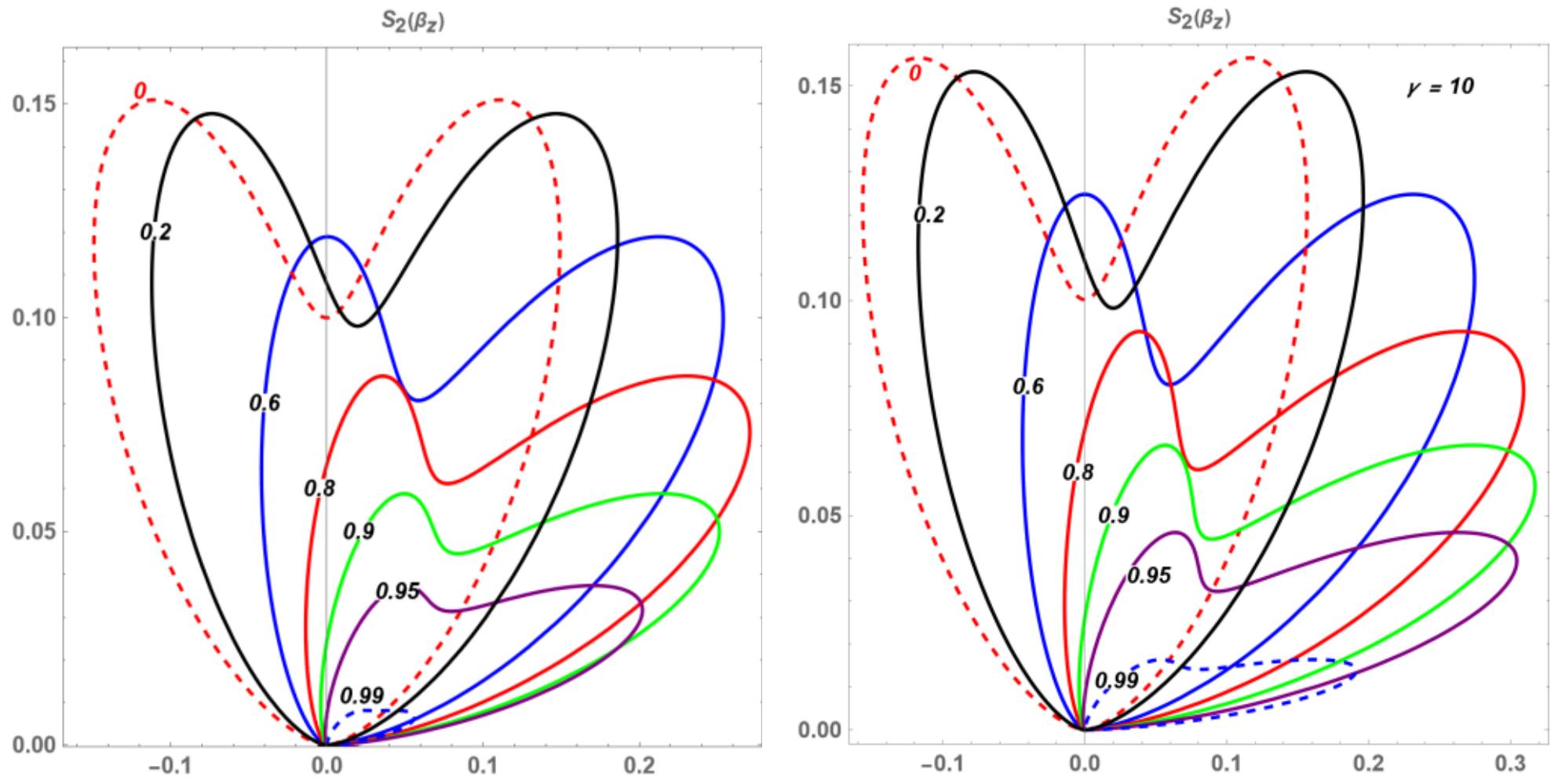

Figure 9. Angular Modal distributions of radiated energy densities normalized to $q^2/2\varepsilon_0 c$ for different values of $\beta_z/\beta_\gamma$ (marked on the graphs by decimal fractions); $\gamma = 5$ (left), $\gamma = 10$ (right); $m = 1$ (top), $m = 2$ (bottom)

The diagrams shown in Fig. 9 and 10 indicate an important feature of the directional properties of the spectral density of radiation, which is not manifested in their integral characteristics (Fig. 8). The amplitudes of the fundamental mode increase with the growth of $\beta_z$, begins to decrease from a certain value, gradually degenerating to the case of linear motion of a particle. The amplitude maximum is achieved at $\beta_z \approx 0.8\beta_\gamma$ for $\gamma = 5$ and at $\beta_z \approx 0.9\beta_\gamma$ for $\gamma = 10$, i.e., at higher energies the degeneracy process begins with larger values of $\beta_z$. The same phenomenon occurs with $m = 2$.

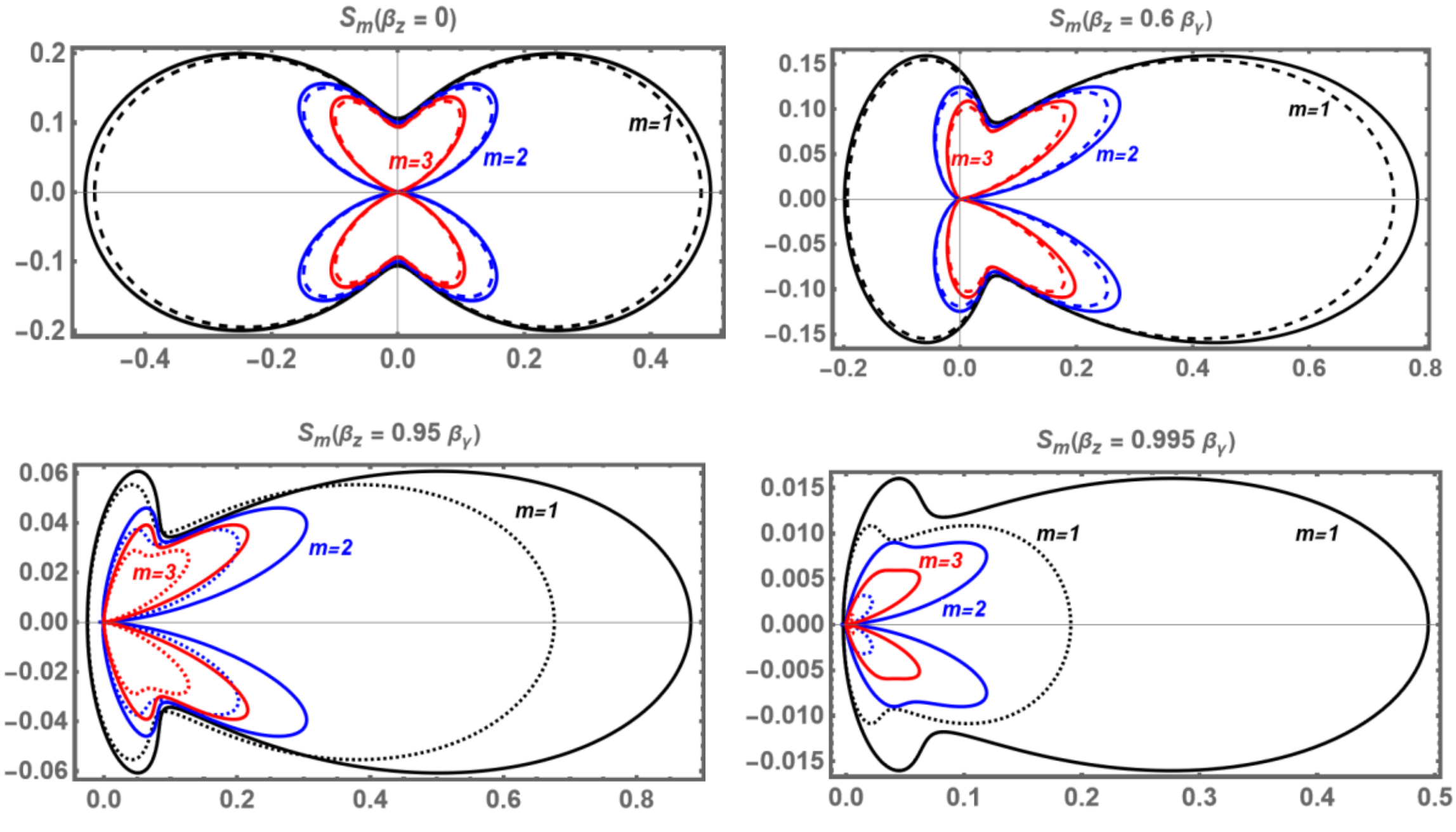

Figure 10. Angular distribution of the normalized modal spectral density of the particle radiation energy for two different values of the particle energy: $\gamma = 5$, (dashed) and $\gamma = 10$ (solid) for several values of $\beta_z$.

As noted earlier and as can be clearly seen from the Figures, only the fundamental mode radiates directly forward ($\theta = 0$). However, as can be seen from the same Figures, radiation from subsequent modes also falls into an arbitrarily small neighbourhood of the direction $\theta = 0$. The comparative contribution of the second mode compared to the contribution of the main mode can be estimated by plotting the dependence of the function $J_1(\beta_z, \theta)/J_0(\beta_z, \theta)$ on the angle $\theta$ for several values of $\beta_z$ (Fig. 11). The horizontal dotted line in the Figure limits the angular region of space in which the relative contribution of the second harmonic does not exceed the level of 0.001 (0.1%). Its intersection points with the curves for different $\beta_z$ determine the angles $\theta_{max}$ at which the radiation pattern should be cut off. The angle values for each of the seven curves are fixed in the Figure. The frequency deviations of the diagram within the spatial cones formed by the given angles are also shown there. As can be seen from the Figure, small values of $\beta_z/\beta_\gamma$ correspond to large angles $\theta_{max}$ and to a small frequency spread $\Delta\tilde{\omega}$, while values of $\beta_z$ close to unity correspond to large angles $\theta_{max}$ and to a large frequency spread $\Delta\tilde{\omega}$.

The normalized frequency spread was calculated using the following equation:

$$\Delta\tilde{\omega} = \frac{1}{1-\beta_z} - \frac{1}{1-\beta_z cos\theta_{max}} \tag{46}$$

Thus, if it is necessary to cover large spatial angles and achieve a high degree of radiation monochromaticity in them, one should use an undulator characterized by a small value of $\beta_z$.

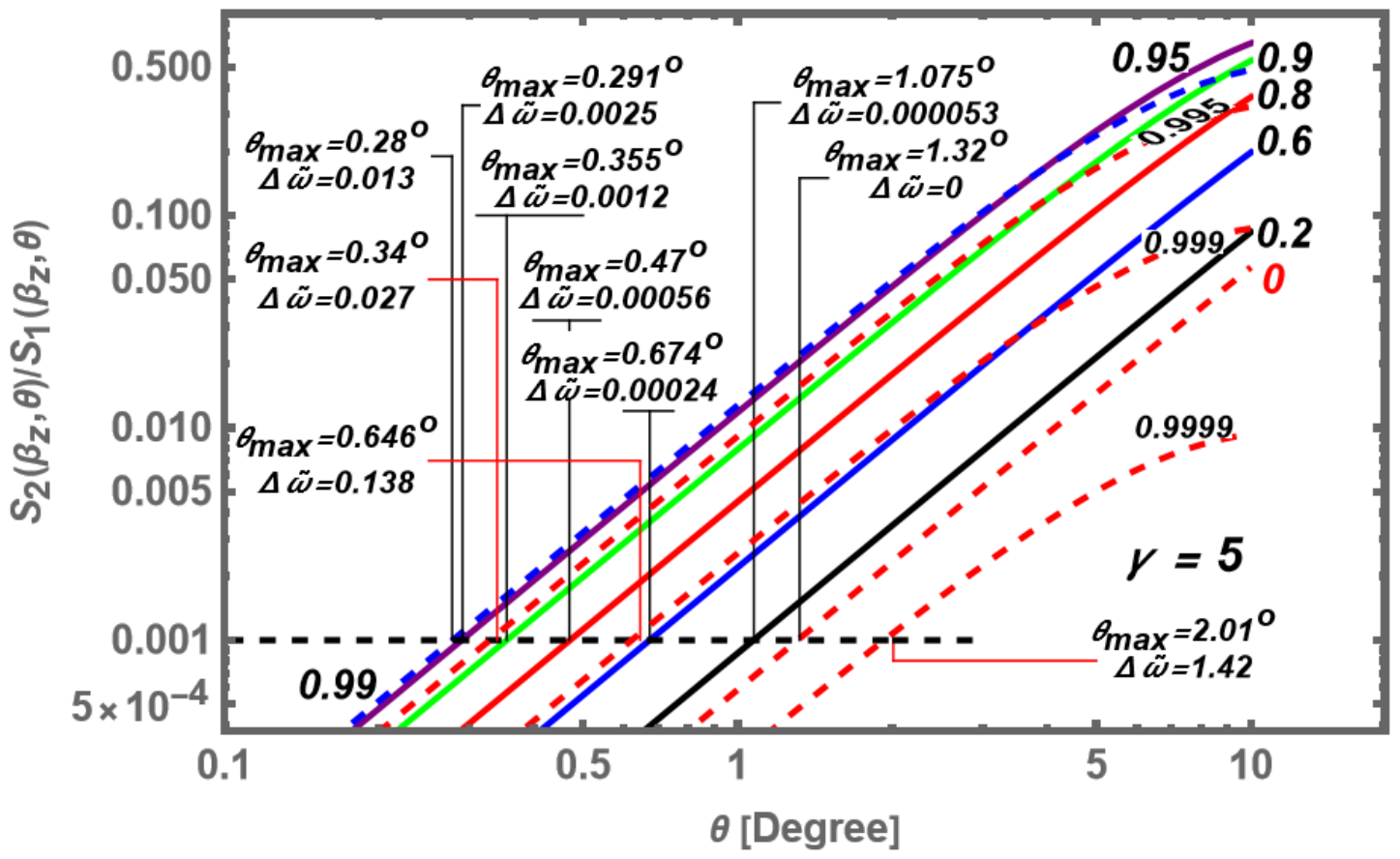

Figure 11. Distributions of functions $S_2(\beta_z)/S_1(\beta_z)$ versus observation angle for different values of $\beta_z$. The numbers near the curves denote the values of $\beta_z/\beta_\gamma$.

## 9. RADIATED POWER

Another important angular modal characteristic of radiation (in addition to the modal spectral density of radiation energy) is the modal power (per revolution) of a particle radiation.

Let's establish a connection between these two characteristics by considering a general problem with an arbitrary source whose radiation power $\tilde{P}$ is represented as an expansion in multipoles $\tilde{P}_m$. Let's assume that the fundamental mode ($m = 1$) radiates at the frequency $\omega_1$ in a certain frequency range $\omega_1^{(1)} < \omega_1 < \omega_1^{(2)}$ and the frequencies of subsequent modes are multiples of m: $\omega_m = m\omega_1$ ($m = 1,2,3, \dots$) and the wavelength of the m-th mode is equal to: $\lambda_m = 2\pi c/m\omega_1$ with $\lambda_m^{(2)} < \lambda_m < \lambda_m^{(1)}$. The energy emitted by $m^{th}$ mode during the wave period is equal to:

$$\tilde{E}_m = \frac{1}{c}\int_{\lambda_m^{(2)}}^{\lambda_m^{(1)}} \tilde{P}_m d\lambda_m = 2\pi \int_{\omega_1^{(1)}}^{\omega_1^{(2)}} \frac{\tilde{P}_m}{m\omega_1^2} d\omega_1 = \int_{\omega_1^{(1)}}^{\omega_1^{(2)}} \bar{P}_m d\omega_1 \tag{47}$$

Here

$$\bar{P}_m = 2\pi \frac{\tilde{P}_m}{m\omega_1^2} \tag{48}$$

is the desired spectral energy density associated with the power $\tilde{P}_m$ of the radiated energy.

In our case $\bar{P}_m = J_m(\omega)$ is defined by the Eqn. 39 and $\omega_1 = \omega_0(1-\beta_z cos\theta)^{-1}$. Thus, the angular distribution of the power emitted by the helical undulator m-th mode can be represented as:

$$\tilde{P}_m(\beta_z, \theta) = \frac{q^2 N m^2 \omega_0^2}{4\pi\varepsilon_0 c} \tilde{S}_m(\beta_z, \theta)$$

$$\tilde{S}_m(\beta_z, \theta) = \frac{(ctg\theta - \beta_z cosec\theta)^2 J_m^2(\tilde{y}_m) + \beta_\varphi^2 J_m'^2(\tilde{y}_m)}{(1-\beta_z cos\theta)^3} \tag{49}$$

Substituting $\beta_z = 0$ into (49) yields the well-known expression for the angular distribution of the power emitted by a particle performing circular motion [15, 16]:

$$\tilde{P}_m^0 = \frac{q^2 m^2 \omega_0^2}{4\pi\varepsilon_0 c}\left\{ctg^2\theta J_m^2(\beta_\gamma m sin\theta) + \beta_\gamma^2 J_m'^2(\beta_\gamma m sin\theta)\right\} \quad (50)$$

The total energy emitted by the m-th mode is determined by integrating Eqn. 49 over the angle $\theta$:

$$\tilde{P}_m(\beta_z) = \frac{q^2 N \omega_0^2}{4\pi\varepsilon_0 c}\tilde{S}_m(\beta_z) \quad (51)$$

with

$$\tilde{S}_m(\beta_z) = m^2 \int_0^\pi \{(ctg\theta - \beta_z cosec\theta)^2 J_m^2(\tilde{y}_m) + \beta_\varphi^2 J_m'^2(\tilde{y}_m)\}(1 - \beta_z cos\theta)^{-3} sin\theta d\theta \quad (52)$$

At $\beta_z = 0$ Eqn. 51 in combination with Eqn. 52 goes over into the well-known equation for the modal radiation power of a particle moving along a circle, which dependence on m has been well studied [15, 16] and, in particular, it has been shown that it has the shape of a hump with a peak corresponding approximately (for very large $\gamma$) to the harmonic number, numerically equal to the cube of the particle's Lorentz factor: $m \sim \gamma^3$. Now, due to the Eqn. 51 it is possible to perform a continuous transition from $\beta_z = 0$ to $\beta_z > 0$ and follow the transformation of the known curve for $\beta_z = 0$ to the forms that describe it for arbitrary values of $\beta_z$ (Fig. 12). As the parameter $\beta_z$ increases, the hump begins to shift towards decreasing harmonic numbers with increasing amplitude. When $\beta_z$ is close to unity, it shifts up to the second harmonic. For $\gamma = 5$ this happens at $\beta_z = 0.97\beta_\gamma$ (Fig. 12, right), and for $\gamma = 10$ - at $\beta_z = 0.99\beta_\gamma$ (Fig. 12, left). With further increase of $\beta_z$, the hump is eliminated and the curve becomes monotonically decreasing with increasing harmonic number. At $\beta_z = 0.98\beta_\gamma$ ($\gamma = 5$) and at $\beta_z = 0.995\beta_\gamma$ ($\gamma = 10$) the amplitude of the curve reaches its maximum value. With further growth of $\beta_z$ ($\beta_z \to 1$), a rapid decline in the overall level of radiation power is observed (see curves for $\beta_z = 0.99\beta_\gamma$ and $\beta_z = 0.995\beta_\gamma$ at $\gamma = 5$ and curves for $\beta_z = 0.998\beta_\gamma$ and $\beta_z = 0.999\beta_\gamma$ at $\gamma = 10$). These curves remain monotonically decreasing with a maximum amplitude at $m = 1$.

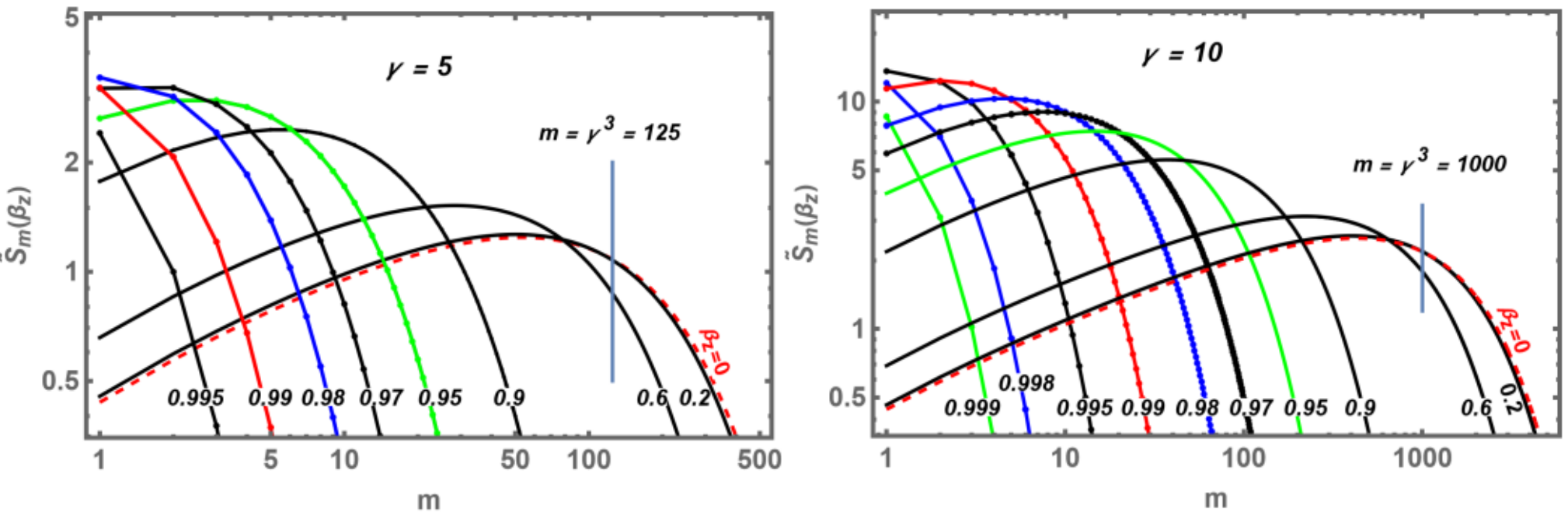

Figure 12. Normalized distributions of the modal radiation power of a particle moving along a spiral trajectory, depending on the mode number; $\gamma = 5$ (right), $\gamma = 10$ (left). The numbers marking the curves on the graphs correspond to values of $\beta_z / \beta_\gamma$.

The integral characteristics shown in Fig. 12, describing the distribution of radiation power by modes, are supplemented by angular diagrams of radiation power for the first two main modes ($m = 1,2$, presented on Fig. 13). The angular diagrams are given for two different energy values: $\gamma = 5$ and $\gamma = 10$ for relatively large values of $\beta_z$: $\beta_z \geq 0.9$ for $\gamma = 5$ and $\beta_z \geq 0.95$ for $\gamma = 10$. The dashed curves show diagrams with those $\beta_z$ whose integral characteristics versus $m$ in Fig. 12 have a hump (the maximum of the distribution falls on the harmonic with a number m greater than unity). As can be seen from Fig. 12, with the growth of $\beta_z$ the level of the hump increases, and, more importantly, the integral contribution of the fundamental mode increases, and, as can be seen from Fig. 13, is accompanied by an increase in its amplitude in the main direction $\vartheta = 0$. With further growth of $\beta_z$, the humped curves smoothly transition into humpless ones (which corresponds to the solid curves in Fig. 13), while both the integral contribution of the fundamental harmonic (Fig. 12) and the amplitude of the fundamental mode in the main direction increase (Fig. 13, top), reaching their maximum at practically the same $\beta_z$, close to unity. Further growth of $\beta_z$ leads to a decrease in the radiated power falling on the first mode and to a decrease in its amplitude. Thus, from the point of view of the maximum share of power falling on the first mode and, at the same time, the maximum value of its amplitude, the preferred options are $\beta_z = 0.98$ and $\beta_z = 0.99$ for $\gamma = 5$ and $\beta_z = 0.995$ and $\beta_z = 0.998$ for $\gamma = 10$.

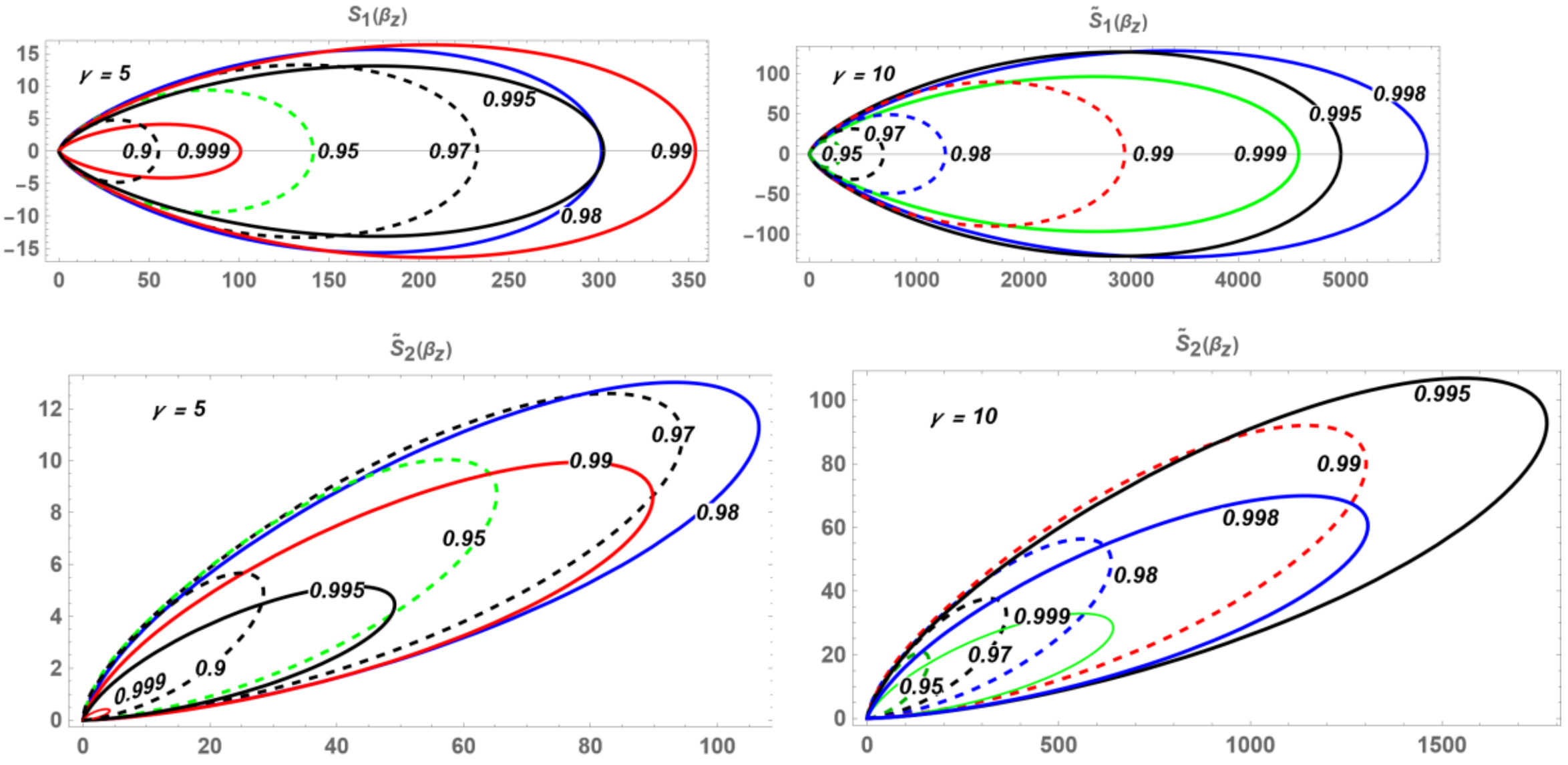

Figure 13. Angular Modal distributions of radiated power normalized to $q^2/2\varepsilon_0 c$ for different values of $\beta_z/\beta_\gamma$ (marked on the graphs by decimal fractions); $\gamma = 5$ (left); $\gamma = 10$ (right); $m = 1$ (top), $m = 2$ (bottom)

One of the criteria for assessing the second harmonic (Fig. 13, bottom) can be considered the angular distance of the maximum of its diagram lobe from the main direction ($\theta = 0$). At $\gamma = 5$ the deviation angle is greatest for $\beta_z = 0.98$ and $\beta_z = 0.97$, and at $\gamma = 10$ for $\beta_z = 0.995$ and $\beta_z = 0.99$.

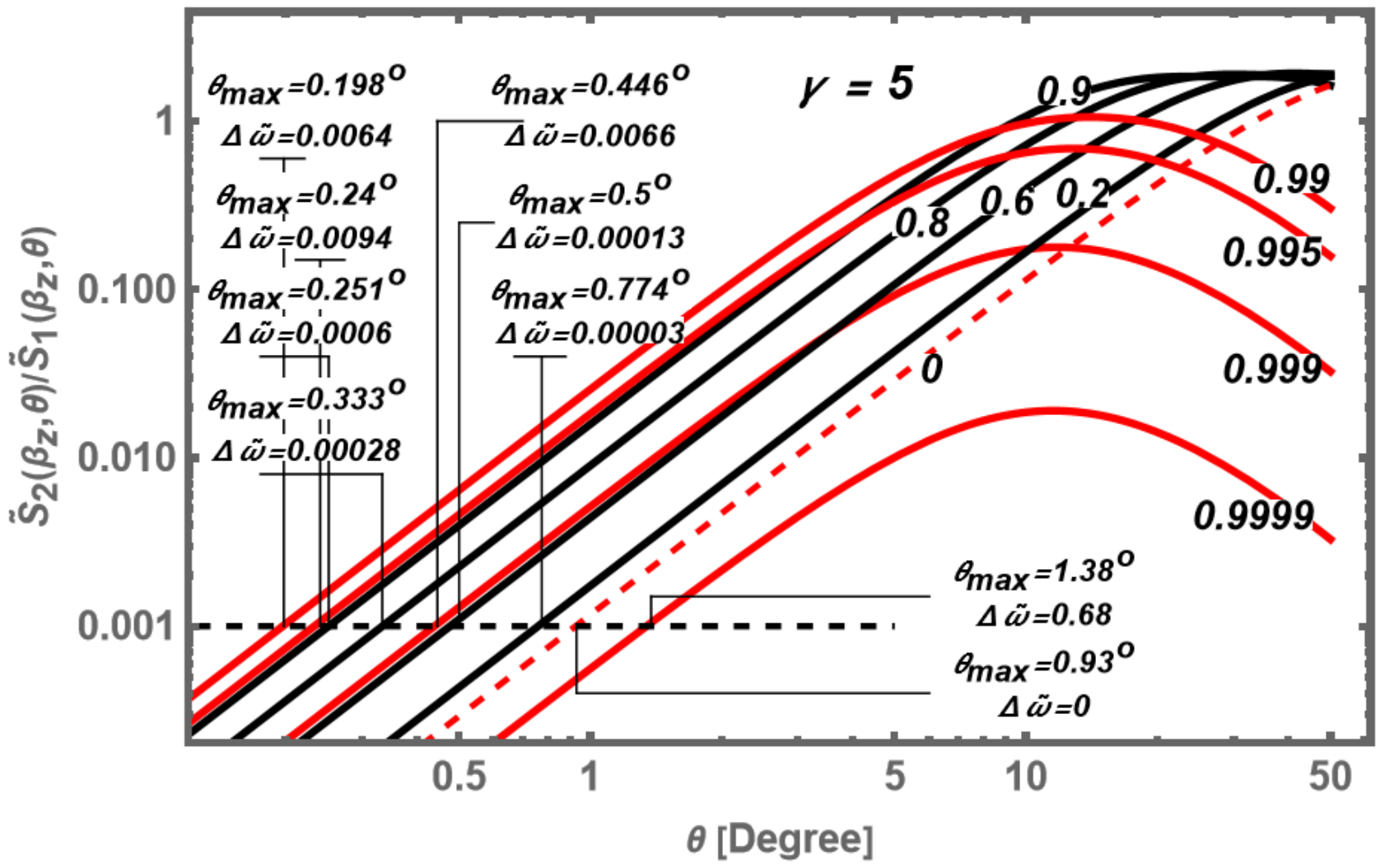

Figure 14. Distributions of functions $\tilde{S}_2(\beta_z)/\tilde{S}_1(\beta_z)$ versus observation angle for different values of $\beta_z$. The numbers near the curves denote the values of $\beta_z$ normalized to $\beta_\gamma$: $\beta_z/\beta_\gamma$.

To estimate the destructive contribution of the second harmonic in the vicinity of the undulator axis, one should, similar to Fig. 11, construct the distributions of the ratio of the functions $\tilde{S}_2(\beta_z,\theta)$ and $\tilde{S}_1(\beta_z,\theta)$ relative to the observation angle $\theta$ for different values of $\beta_z$ (Fig. 14). In Figure 14, the red curves are plotted for $\beta_z < 0.98\beta_\gamma, \gamma = 5$ with humped modal distributions of radiated power (Fig. 12, top), while the black curves refer to humpless distributions.

The growth of $\beta_z$, starting from $\beta_z = 0$, which corresponds to the movement of a particle along a closed circle, is initially accompanied by an increase in the level of humpless curves up to the peak level of transition from humpless curves to humped ones. Further growth of $\beta_z$ is accompanied by a decrease in the levels of the lines $\tilde{S}_2(\beta_z,\theta)/\tilde{S}_1(\beta_z,\theta)$, corresponding to the humped curves. The level of each curve determines the value of the spatial angle within which the share of the contribution of the second harmonic (compared to the first ones) does not exceed a given value. The intersection points of the horizontal dashed line segment with the inclined curves of the graphs of the function $\tilde{S}_2(\beta_z,\theta)/\tilde{S}_1(\beta_z,\theta)$ determine the spatial angles $\theta_{max}$ within which the value of $\tilde{S}_2(\beta_z,\theta)/\tilde{S}_1(\beta_z,\theta)$ does not exceed the value of 0.001 (0.1%). As can be seen from Fig. 14, the mentioned angle has its smallest values for $\beta_z$, corresponding to the transition region from humpless distributions to humped ones (in this region, both the integral radiated power and the power radiated in the main direction have the greatest value). As we move away from the transition region, this angle increases. Thus, the largest angles correspond to both very small values of $\beta_z$ (humped distributions of total modal power in Fig. 12) and when its values very close to unity (humpless distributions of total modal power in Fig. 12).

## 10. OPTIMIZATION PROBLEMS

As follows from the foregoing, the main characteristics of radiation are modal spectral radiation density (39) and modal radiation power (49). Their modal integral characteristics (43) and (52) versus mode number $m$ for $\gamma = 5$ and $\gamma = 10$ are presented in the form of sequences of curves corresponding to certain values of $\beta_z/\beta_\gamma$ in Fig. 8 and 12. Using Eqn. 39 and 49, their angular diagrams for the fundamental ($m = 1$) and subsequent ($m = 2$) modes are also calculated (Fig. 9 and 13) and their superpositions on each other are given for $\gamma = 5$ and $\gamma = 10$ (Fig. 10), which allows one to form an idea of the spatial distribution of radiation in the far zone.

Since the main interest is in the properties of the fundamental mode radiation in the main direction and in its vicinity, its angular and frequency characteristics in this direction were investigated. In particular, for different values of $\beta_z/\beta_\gamma$, the spatial angle around the main direction was estimated, within which the contribution of the second harmonic does not exceed a certain level. These estimates are given in Fig. 11 and 14. The values of the frequency spread of radiation inside the spatial cone corresponding to the determined angles are also given there.

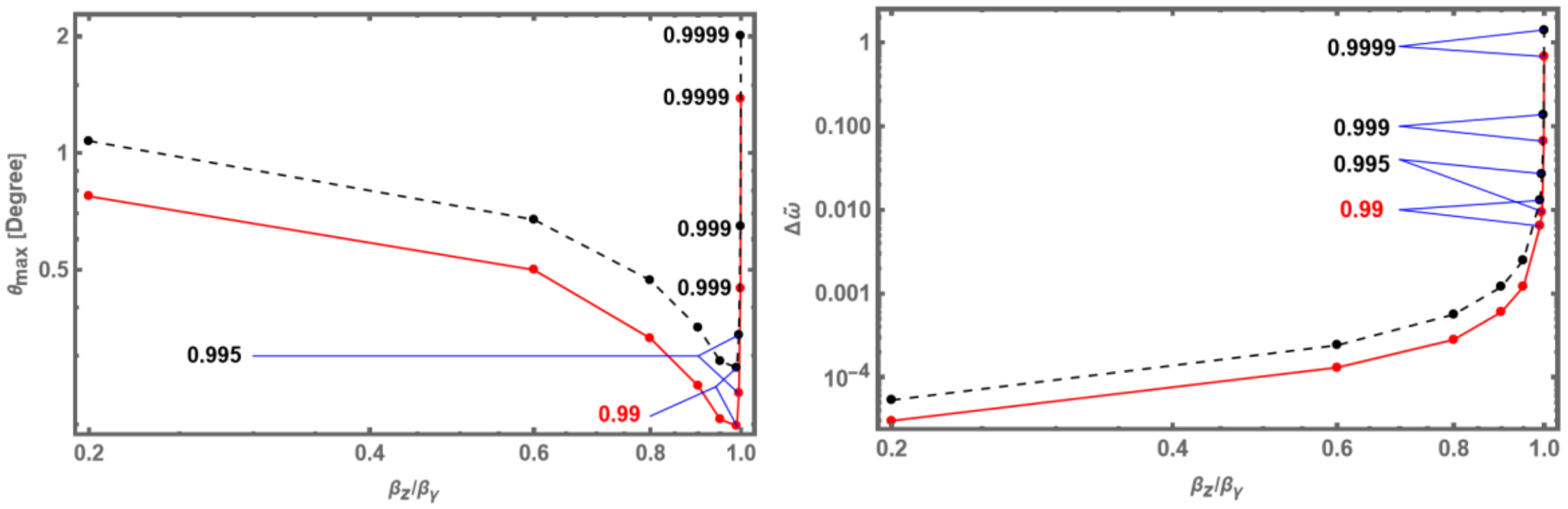

Figure 15. Comparison of angles $\theta_{max}$ (left) and frequency bands $\Delta\omega$ (right) for the modal spectral energy density (black dotted line, taken from Fig. 11) and modal radiated power (red solid line, taken from Fig. 14); $\gamma = 5$.

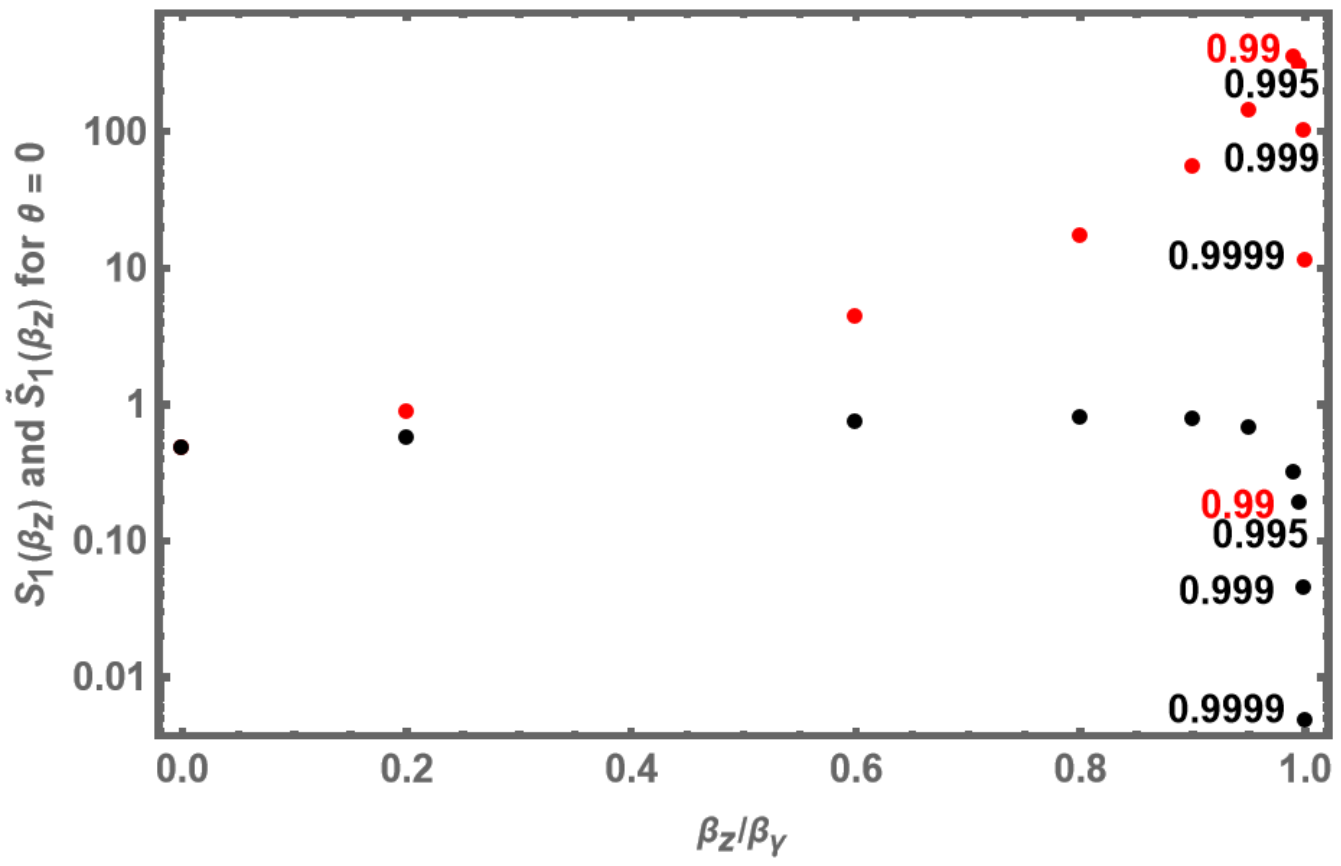

Figure 16. Comparison of fundamental mode amplitudes for the spectral energy density (black, taken from Fig. 9, top, left) and modal radiated power (red, taken from Fig. 13, top, left); $\gamma = 5$.

The information for the case of $\gamma = 5$, contained in Figures 8-14 in a form convenient for comparison and research is presented in Figures 15 and 16. From Fig. 15, in particular, it follows that the curves of the dependencies for $\vartheta_{max}$ (Fig. 15, left) and for $\Delta\tilde{\omega}$ (Fig. 15, right) for the spectral density of the radiated energy and for the radiated power with some shift almost exactly repeat each other. It is noticeable that for both characteristics, small values of $\beta_z/\beta_\gamma$ correspond to comparatively large values of angles $\vartheta_{max}$ with a simultaneous small relative frequency spread $\Delta\tilde{\omega}$. So, if it is necessary to obtain a large angular range with a controllably small destructive contribution from the second harmonic, one can use small $\beta_z$-characteristics without taking into account the small amplitude value both for the radiated power and for the spectral energy density (Fig. 16). The growth of the parameter $\beta_z$ leads to a decrease in the angle $\vartheta_{max}$. Its minimum value is reached at $\beta_z/\beta_\gamma = 0.99$, then its sharp growth occurs, accompanied by an equally sharp increase in the magnitude of the frequency spread.

The amplitude curve for the modal radiation power (Fig. 16, red) increases monotonically with the growth of $\beta_z/\beta_\gamma$, and reaches a maximum at the same value of $\beta_z/\beta_\gamma = 0.99$, at which the angular curves (Fig. 15, left) reach a minimum level. With further increase in $\beta_z/\beta_\gamma$, it experiences a sharp decline.

The amplitude curve for the modal spectral density of the radiated energy is a slowly increasing function up to $\beta_z/\beta_\gamma = 0.8$, then it begins to decrease slowly, and at $\beta_z/\beta_\gamma > 0.95$ it experiences a sharp decline.

Thus, the maximum amplitude is accompanied by a minimum value of the angle $\theta_{max} = 0.198^o$ and an acceptable level of frequency spread $\Delta\widetilde{\omega}\sim 0.01$, and large angles $\theta_{max}$ correspond to small frequency spreads, but, unfortunately, also small amplitudes. It is not possible to achieve absolute optimization in all three parameters, but compromise solutions are available: large angles $\theta_{max}$ and a small frequency spread $\Delta\widetilde{\omega}$ with low radiation intensity in the main direction or high radiation intensity at small angles $\theta_{max}$ and a relatively low degree of its monochromaticity. The final choice depends on the goals of the experiment or the functional needs of the device of which the undulator is to serve.

## 11. RESUME

The method developed in this article is based on an original technique, which consists of comparing the exact solution, constructed in the form of an expansion in cylindrical multipoles, but containing an undetermined coefficient $\chi_m$ (24), with an approximate solution (18), in order to determine this coefficient. The uniqueness of the method used lies in the demonstrated possibility of determining the exact value of this coefficient by comparing the far-zone and short-wave asymptotic of the exact solution with its approximate analogue (18), which made it possible to construct an exact solution for the undulator radiation field and made it possible to compare the exact formula for the spectral density of radiation energy with a well-known approximate formula and to evaluate the applicability criteria of the latter. The permitted frequency ranges are specified and the possibility of discretizing the distribution of the spectral density of radiation energy is shown.

The presence of an exact solution made it possible to construct explicit expressions for the Poynting vector, derive exact formulas for the Doppler effect, and identify the frequency regions responsible for forward and backward radiation for each term of the multipole expansion.

A connection has been established between the formulas describing the spectral density of radiation energy and the radiation power for the terms of the multipole expansion.

The possibility of a continuous transition from the modal distribution of the spectral density of radiation energy and radiated power of a particle moving along a spiral trajectory to the same indicators when it moves in a closed circle.

The analysis of the properties of spatial distributions of radiation made it possible to develop quantitative criteria for assessing the qualitative properties of radiation and to evaluate the possibility of its optimization according to the main parameters (the degree of monochromaticity of radiation in a selected angular cone, the magnitude of the longitudinal component of the particle velocity $\beta_z$ and the amplitude characteristics of the radiation).

ACKNOWLEDGMENTS

The work was supported by the Science Committee of RA, in the frames of the research projects № 21T-1C239 and № 23SC-CNR-1C006.